\documentclass[namedreferences]{solarphysics}
\usepackage[hyperref,optionalrh,solaromanenum]{spr-sola-addons} 
\usepackage{graphicx}                    
\usepackage{amssymb}                    
\usepackage{color}                       
\usepackage{breakurl}                         
\usepackage{pdflscape}
\hypersetup{colorlinks, citecolor=blue, filecolor=black, linkcolor=black, urlcolor=black}

\newcommand{\etal}{{\it et al.}}
\newcommand{\degree}{\hbox{$^{\circ}$}}

\chardef\us=`\_

\begin{document}

\begin{article}

\begin{opening}

\title{Active Region Photospheric Magnetic Properties Derived from Line-of-sight and Radial Fields}

%
\author[addressref={aff1},email={jordan.a.guerra@tcd.ie}]{\inits{J.A.}\fnm{J.A.}~\lnm{Guerra}}\orcid{0000-0001-8819-9648}
\author[addressref={aff1},email={sunpark@tcd.ie}]{\inits{S.-H.}\fnm{S.-H.}~\lnm{Park}}\orcid{0000-0001-9149-6547}
\author[addressref={aff1},email={peter.gallagher@tcd.ie}]{\inits{P.T.}\fnm{P.T.}~\lnm{Gallagher}}\orcid{0000-0001-9745-0400}
\author[addressref={aff2},email={jkonto@noa.gr}]{\inits{I.}\fnm{I.}~\lnm{Kontogiannis}}\orcid{0000-0002-3694-4527}
\author[addressref={aff2},email={manolis.georgoulis@academyofathens.gr}]{\inits{M.K.}\fnm{M.K.}~\lnm{Georgoulis}}\orcid{0000-0001-6913-1330}
\author[addressref={aff3},email={shaun.bloomfield@northumbria.ac.uk}]{\inits{D.S.}\fnm{D.S.}~\lnm{Bloomfield}}\orcid{0000-0002-4183-9895} 

%
\runningauthor{J.A.~Guerra \etal}
\runningtitle{Active Region Properties from LOS-/Radial-magnetograms}

\address[id={aff1}]{School of Physics, Trinity College Dublin, College Green, Dublin 2, Ireland}
\address[id={aff2}]{Research Center for Astronomy and Applied Mathematics (RCAAM), Academy of Athens, 4 Soranou Efesiou Street, Athens, GR-11527, Greece}
\address[id={aff3}]{Northumbria University, Newcastle upon Tyne, NE1~8ST, UK}

\begin{abstract}
The effect of using two representations of the normal-to-surface magnetic field to calculate photospheric measures that are related to active region (AR) potential for flaring is presented. Several AR properties were computed using line-of-sight ($B_{\rm los}$) and spherical-radial ($B_{r}$) magnetograms from the Space-weather HMI Active Region Patch (SHARP) products of the \emph{Solar Dynamics Observatory}, characterizing the presence and features of magnetic polarity inversion lines, fractality, and magnetic connectivity of the AR photospheric field. The data analyzed corresponds to $\approx$\,4,000 AR observations, achieved by randomly selecting 25\% of days between September 2012 and May 2016 for analysis at 6-hr cadence. Results from this statistical study include: 
 i) the $B_{r}$ component results in a slight upwards shift of property values in a manner consistent with a field-strength underestimation by the $B_{\rm los}$ component; 
 ii) using the $B_{r}$ component results in significantly lower inter-property correlation in one-third of the cases, implying more independent information about the state of the AR photospheric magnetic field; 
 iii) flaring rates for each property vary between the field components in a manner consistent with the differences in property-value ranges resulting from the components; 
 iv) flaring rates generally increase for higher values of properties, except Fourier spectral power index that has flare rates peaking around a value of $5/3$. 
 These findings indicate that there may be advantages in using $B_{r}$ rather than $B_{\rm los}$ in calculating flare-related AR magnetic properties, especially for regions located far from central meridian.
\end{abstract}

%

\end{opening}

%
%
\section{Introduction}\label{s:int}

It is well known that the energy that powers solar flares and coronal mass ejections (CMEs) is slowly built-up in the magnetically-structured coronae of active regions (ARs) and then quickly released {\it via} the reconfiguration of magnetic field \citep{2017ApJ...836...17A}. In the absence of direct 
 measurements of the coronal magnetic field (or uniquely defined non-potential reconstructions), flare forecasting has relied on photospheric information instead. In particular, flare prediction has given great importance to the normal-to-surface field component, $B_{\rm n}$, for several reasons. First, because the normal component can be assumed from the line-of-sight (LOS) component under certain circumstances ({\it i.e.}, close to disk centre). Second, the normal component provides the positive- and negative-polarity areas in the photosphere where the 3D magnetic field -- that extends into the corona -- is rooted. In this sense, the normal component should provide information about the magnetic energy flux that flows from the photosphere into the coronal volume occupied by the AR.

Previous studies have focused on quantifying spatial patterns of the photospheric field using the LOS component restricted to regions near disk centre. In terms of the physical characteristic or process to be analyzed, these studies can be group into a few categories:

\begin{enumerate}
\item those detecting and parameterizing magnetic polarity inversion lines (MPILs) as proxies of current-carrying magnetic structures emerging through the photosphere \citep[{\it e.g.},][]{2003JGRA..108.1380F,2007ApJ...655L.117S};
\item those quantifying the complex spatial patterns of the photospheric field in terms of the field's multi-scale behavior \citep[{\it e.g.}\,][]{2005ApJ...629.1141A,2008SoPh..248..311H,2015SoPh..290..335G} and fractal properties \citep[{\it e.g.}\,][]{2008SoPh..248..297C,2010ApJ...722..577C,2014SoPh..289.2525E,2005ApJ...631..628M};
\item those characterizing the field connectivity based on photospheric information \citep[{\it e.g.},][]{2007ApJ...661L.109G,ahmed2010,2014ApJ...785...88Z,2006ApJ...646.1303B}.
\end{enumerate}

For ARs away from disk centre, some considerations need to be taken into account when using LOS data. As ARs move away from disk centre, the observing angle, $\theta'$, increases and the contributions of magnetic vector magnitude and direction to the LOS component change \citep{2017SoPh..292...36L}. In addition, the LOS component samples an increasing portion of the parallel-to-surface component, resulting in artifacts like unphysical limb-effect MPILs in magnetic structures such as ARs. The common $\mu$-angle correction, $B_{\rm n} = B_{\rm los}/\mu$ where $\mu=\cos\left(\theta'\right)$, assumes all photospheric field is normal and therefore results in enhancement of such artifacts. Thus, it seems reasonable to assume that a more sophisticated approximation of the normal component of the magnetic field will represent more consistently any property calculated far from disk centre.

Before the launch of \emph{Solar Dynamics Observatory} \citep[SDO;][]{2012SoPh..275....3P}, some statistical studies of flaring-related magnetic properties of ARs employed a more realistic normal component measured from space- and ground-based vector magnetograms \citep[see {\it e.g.}][]{2007ApJ...656.1173L}. These studies contributed greatly to the understanding of flaring active regions, but the regular use of vector-magnetic field in flare forecasting was not practical until the \emph{Heliospheric Magnetic Imager} \citep[HMI;][]{sdo_hmi} on board SDO started providing near-realtime full-disk vector magnetograms. The HMI vector data preparation pipeline provide the $B_{\rm los}$ component from the \textsf{Mharp} series and the surface-normal component in the form of the spherical-radial field, $B_{r}$ from the Milne-Eddington (ME) inversion series. Therefore, a statistical study focusing on the difference between AR properties calculated from $B_{\rm los}$ and $B_{r}$ seems appropriate and timely. In Section~\ref{s:data}, the data sample used and pre-processing methods are described, while Section~\ref{s:properties} describes the AR properties investigated. Section~\ref{s:results} presents the results, focusing on the difference in properties calculated using the $B_{\rm los}$ and $B_{r}$ data (Section~\ref{ss:comp}), their variation with AR longitudinal position (Section~\ref{ss:var_pos}), the correlation between properties (Section~\ref{ss:par_correl}), and their flare association (Section~\ref{ss:flaring}). Finally, Section~\ref{s:concl} presents concluding remarks.

\section{Data Sample}\label{s:data}
The AR properties included in this study were calculated from the Space-weather HMI Active Region Patch \citep[SHARP;][]{2014SoPh..289.3549B} data products from SDO. SHARP data products contain AR magnetograms derived from the HMI full-disk magnetograms. Vector and LOS maps are derived from the \textsf{hmi.ME\_720s\_fd10} and \textsf{hmi.Mharp\_720s} series, respectively. ARs are detected as a group of pixels exceeding a threshold value of 100 G in the unsigned LOS field. These group of strong field pixels are then separated into individual HMI Active Region Patches \citep[HARPs;][]{2014SoPh..289.3483H}, each uniquely identified by a HARP number and the observation time. SHARP data contain, amongst other information, the disambiguated photospheric magnetic field in two sampling systems: field magnitude, azimuth and inclination angles (with respect to the LOS) in native image-plane sampling; spherical radial, longitudinal and latitudinal components of the field in cylindrical equal area (CEA) deprojected sampling. The near-realtime (NRT) version of SHARP data is created and made available with short delay from the observation time. In this version the selected HARP field-of-view (FOV) varies according to the AR evolution.

\begin{figure}[!t]
 \centerline{\includegraphics[width=0.5\textwidth,clip=]{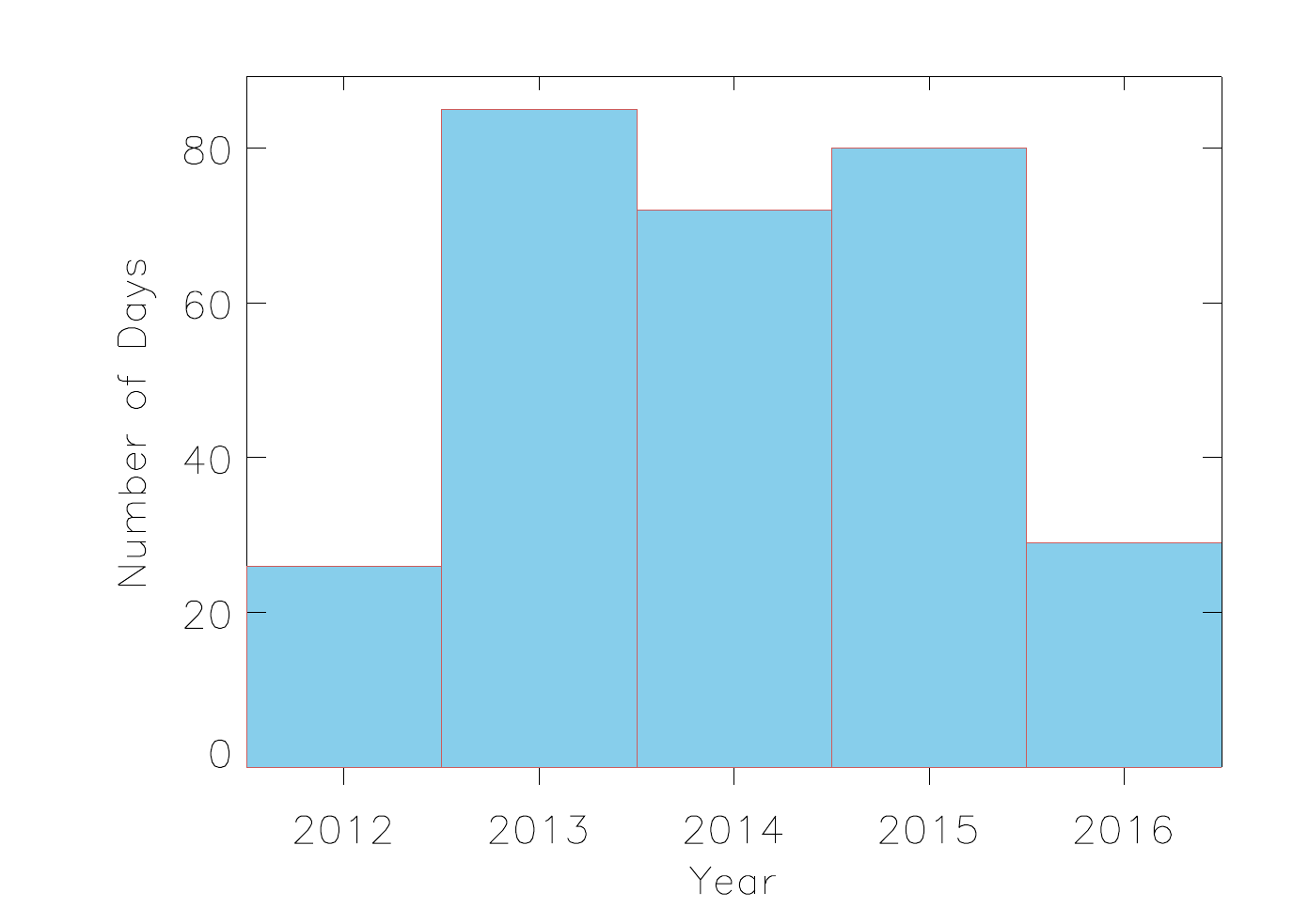}\includegraphics[width=0.5\textwidth,clip=]{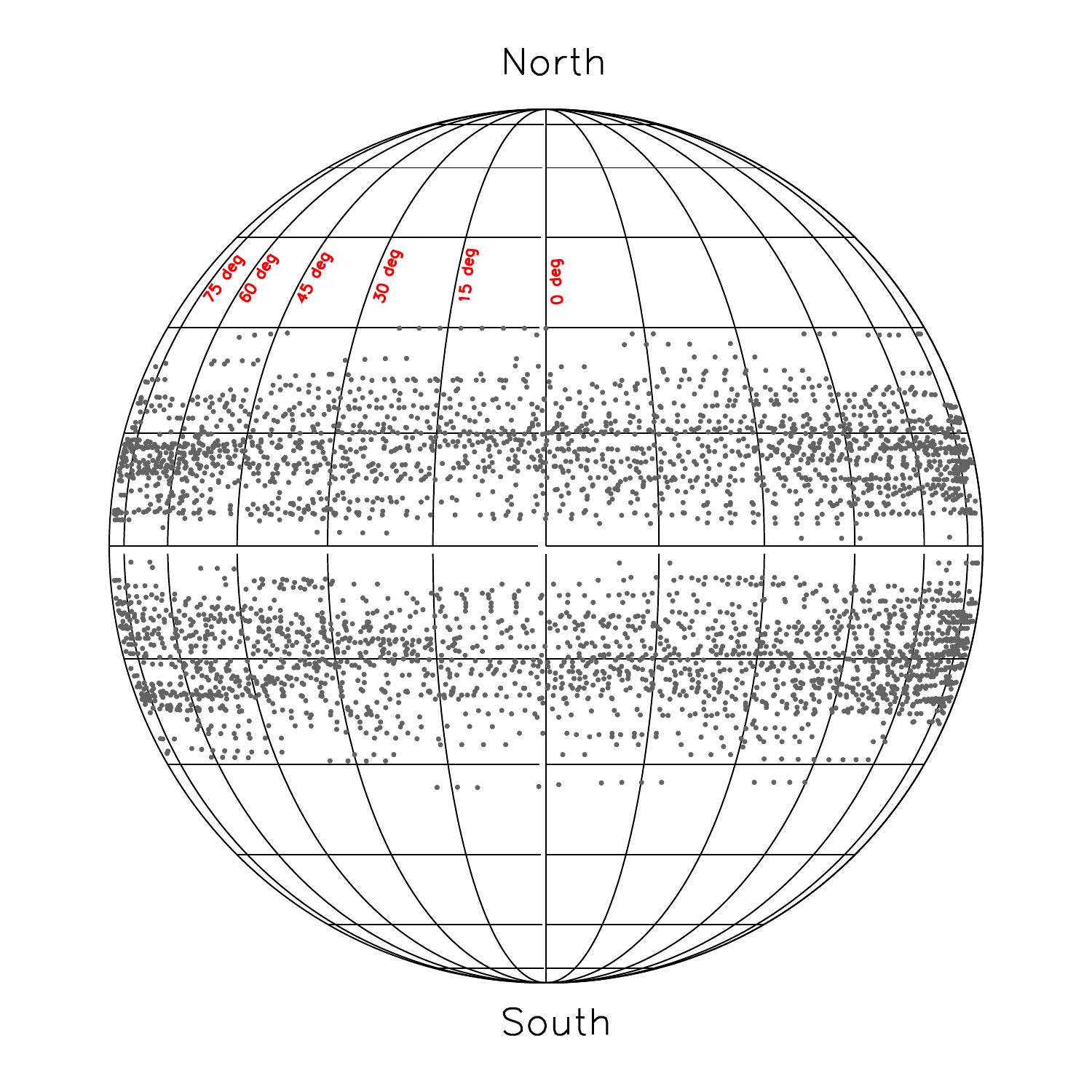}}
 \caption{Graphic representation of the of the data sample analyzed. Sample corresponds to 25\% of days randomly selected between 15 September 2012 and 17 May 2016. Distribution of the selected days by year (left panel) and solar disk locations of SHARPs (right panel).}
 \label{fig:sample}
 \end{figure}

For this study, CEA NRT data were accessed from the \textsf{hmi.sharp\_cea\_720s\_nrt} series. A robust sample of the entire SHARP repository was selected to perform a significant statistical study of flare-related AR properties. This sample corresponds to 25\% of days between 15 September 2012 and 17 May 2016, selected from a random uniform distribution. The distribution of selected days is displayed in the left panel of Figure \ref{fig:sample}. For each selected day, four time stamps are analyzed that correspond to a cadence of 6\,hr ({\it i.e.} times closest to 00:00, 06:00, 12:00, and 18:00\,UT). On average there are approximately ten SHARP regions on disk at any time in the sample. Thus 12,733 SHARPs were analyzed, with the right panel of Figure \ref{fig:sample} displaying the spatial distribution of heliographic (HG) coordinates ($\phi$ and $\theta$ for longitude and latitude, respectively). It can be seen that SHARP locations are distributed uniformly over the two hemispheric active latitude bands and across all longitudes.

%

Data quality selection is done before calculating the properties. In the pre-processing stage, first NRT SHARP data is checked for bad-quality images (images containing mostly null values), missing header information, or failing of the world coordinate system \citep[WCS;][]{2006A&A...449..791T} calculation. If any of these problems are encountered, the HARP is considered as bad data and no properties are calculated from it. HARP FOVs can contain off-limb pixels when the patches are detected near the solar limb. Thus, HARPs are pre-processed in order to exclude off-limb pixels from subsequent calculations. An edge-detection procedure was implemented to detect the solar limb in CEA de-projected images. First, the gradient of the image is calculated by convolving the image with a Sobel filter \citep[see {\it e.g.}][]{2015A&A...579A..64A}. Then using the WCS, the HG location can be assigned to each pixel on the limb. Finally, the lowest value of longitudinal distance is selected as the new maximum longitude for the HARP FOV. The result of this process is an image trimmed in the horizontal ({\it i.e.} longitudinal) direction.

\section{Active Region Magnetic Properties}\label{s:properties}

The quantities considered here correspond to AR magnetic field properties for most of which significance in flare prediction has been  shown \citep[see {\it e.g.}][]{2008ApJ...688L.107B}. Each feature described below can provide one or more properties to the study. However, properties presented here correspond to a subgroup selected for the purpose of illustrating differences between $B_{\rm los}$- and $B_{r}$-calculated properties. This subgroup covers different aspects of ARs, from photospheric features of interest such as MPILs, to the entire AR flux distribution.

\subsection{Magnetic Polarity Inversion Line (MPIL) Properties}\label{ss:pils}

MPILs, or neutral lines, in the photosphere of ARs separate distinct patches of positive and negative flux and several studies have related flare occurrence to MPIL properties \citep[{\it e.g.}\,][]{2007ApJ...655L.117S,2010ApJ...723..634M,2012ApJ...757...32F}. A subset of MPILs have also been identified as strong MPILs (or $^{\star}$MPILs) based on:
\begin{enumerate}
\item strong horizontal gradients in the vertical-field component across the MPIL;
\item strong horizontal-field component over the MPIL.
\end{enumerate} 

$^{\star}$MPILs can be considered as photospheric evidence of the emergence of highly twisted fields and the place in ARs where flux cancellation is likely to take place. \citet{Bokenkampthesis} developed an algorithm to define $^{\star}$MPILs using a three-stage process based on the original $^{\star}$MPIL detection routine of \citet{2003JGRA..108.1380F}. First, MPILs are identified as contours of zero field in a strongly-smoothed $B_{\rm n}$. Then a vector-magnetic field map is calculated from the smoothed $B_{\rm n}$ image using a potential-field model \citep[{\it e.g.}\,][]{1981A&A...100..197A}. $^{\star}$MPILs are then extracted from the detected MPILs by thresholding: i) the gradient of the vertical magnetic field; ii) the strength of the horizontal magnetic field in the potential-field model. Second, the same process of detection is repeated for a less-smoothed $B_{\rm n}$ image. $^{\star}$MPILs are determined by comparing between $^{\star}$MPILs$_{1}$ and $^{\star}$MPILs$_{2}$ identified in the first and second stages, respectively; retained $^{\star}$MPILs correspond to portions of $^{\star}$MPILs$_{2}$ that are separated from $^{\star}$MPILs$_{1}$ by distances less than a specified threshold of 11\,Mm. Finally, this comparison method is applied between $^{\star}$MPILs identified from an unsmoothed image and the newly identified $^{\star}$MPILs with the same distance threshold of 11\,Mm. Note that in ARs with complicated magnetic configurations, the $^{\star}$MPIL detection algorithm usually identifies multiple $^{\star}$MPILs within one magnetogram FOV.

In this study, a modification of the \citet{Bokenkampthesis} detection algorithm was implemented by skipping the final stage in the three-stage process to reduce computation time. Comparison between original- and modified-algorithm results show very little difference between detected $^{\star}$MPILs. Therefore, the modified two-stage process is used here to find $^{\star}$MPILs using $B_{\rm n}=B_{r}$ and $B_{\rm n} =$ $\mu-$corrected $B_{\rm los}$, with smoothing factors of 6 and 3 pixels used in the first and second stages of the calculation, respectively. Flux threshold values used to define the strong segments of MPILs are 120 and 100 Mx cm$^{-2}$ in each stage, correspondingly. From the determined $^{\star}$MPILs, several properties can be obtained such as maximum and total lengths of $^{\star}$MPIL segments ($L_{\rm max}$ and $L_{\rm tot}$, respectively), and the total unsigned flux $\Phi_{\rm MPIL}$ near $^{\star}$MPILs. In this paper, only the total length is reported.

\subsection{Decay Index}\label{ss:decayi}

The horizontal field decay index, $n_{\rm hor}$, is a quantity related to the onset of the torus instability for current-carrying flux ropes in ARs. The torus instability has been extensively studied in relation to the success or failure of ARs in producing CMEs \citep[{\it e.g.}\,][]{2008ApJ...679L.151L, 2014ApJ...785...88Z,2015ApJ...814..126Z}. However, a statistical study of how the decay index relates to flare occurrence has not been conducted. In order to determine $n_{\rm hor}$, a 3D vector-magnetic field is required to be derived from the photospheric field using a field extrapolation. In the coronal volume of an AR, $n_{\rm hor}$ is defined as the localized gradient of the logarithmic horizontal magnetic field with logarithmic height,
\begin{equation}
n_{\rm hor} = - \frac{{\rm d}\log\left(B_{\rm hor}\right)}{{\rm d}\log\left(h\right)} \ ,
\end{equation}
where $B_{\rm hor}$ is the horizontal component of the magnetic field and $h$ is height above the photosphere. In order to derive the 3D vector-magnetic field above the AR, a potential-field model is implemented \citep[{\it e.g.}\,][]{1981A&A...100..197A}, using $B_{r}$ or $\mu-$corrected $B_{\rm los}$ as photospheric lower boundary input. After a 3D array of localized $n_{\rm hor}$ values is determined from this model, several $^{\star}$MPIL-related decay index properties can be derived. Included here are: i) minimum height of critical decay index ($n_{\rm cr} = 1.5$) above $^{\star}$MPILs, $h_{\rm min}$; ii) maximum ratio of $^{\star}$MPIL length to $h_{\rm min}$ from each $^{\star}$MPIL in an AR, $\left(L/h_{\rm min}\right)_{\rm max}$. A decay index calculation based on a potential field provides an estimate of how quickly the horizontal component of the field due to sources ({\it e.g.} overlying arcade fields in active regions), external to a current-carrying non-potential field, decreases with height. There have been several studies \citep{2005ApJ...630L..97T,2006PhRvL..96y5002K,2007ApJ...668.1232F,2008SoPh..249...75L} establishing the gradient of a magnetic field overlying an erupted flux rope as important to understanding kink and torus instabilities for solar eruptions.

\subsection{Schrijver's \textit{R} value}\label{ss:r_val}

The $R$ value quantifies the unsigned photospheric magnetic flux near ({\it i.e.} within $\approx$\,15\,Mm of) strong-field high-gradient MPILs within active regions. This property and its usefulness in forecasting was first investigated by \citet{2007ApJ...655L.117S}. The presence of such MPILs indicates that twisted magnetic structures which carry electrical currents (associated with the potential for flare activity) have emerged into the active region through the solar surface. Therefore, $R$ represents a proxy for the free magnetic energy that is available for release in a flare.

The algorithm for calculating $R$ is relatively simple, computationally inexpensive, and was developed using $B_{\rm los}$  magnetograms from the \emph{Michelson Doppler Imager} \citep[MDI;][]{1995SoPh..162..129S} onboard the \emph{Solar and Heliospheric Observatory} \citep[SOHO:][]{Domingo:1995}. First, a bitmap is constructed for each polarity in a magnetogram, indicating where the absolute magnitude of the magnetic flux density exceeds the threshold value of 150\,Mx\,cm$^{-2}$. These bitmaps are then dilated by a square kernel of $3 \times 3$\,pixels and the areas where the bitmaps overlap are defined as strong-field MPILs. This combined bitmap is then convolved with a Gaussian filter of 15\,Mm FWHM, which corresponds to the peak value of the distribution of flare ribbons distance (as observed in EUV images) from MPILs (see \citet{2007ApJ...655L.117S} for further detail). Finally, the Gaussian-convolved bitmap is multiplied with the absolute flux value of the $B_{\rm los}$ magnetogram and $R$ is calculated as the sum over all pixels.

For calculating $R$ values from the magnetogram sample, it is necessary to pre-process the data and implement some changes in the corresponding algorithms. First, SHARP magnetograms are resampled from 0.03$^{o}$\,pixel$^{-1}$ to 0.12$^{o}$\,pixel$^{-1}$ ({\it i.e.} equivalent to MDI resolution in a CEA deprojection) to match the definition in \citet{2007ApJ...655L.117S}. Second, to calculate $R$ using the $B_{r}$ component it is necessary to use a different (and higher) threshold of 300\,Mx\,cm$^{-2}$ in defining strong-field MPILs. This is because the $B_{r}$ component has increasing levels of noise at increasing distances from disk centre, due to larger contributions from the linear Stokes polarization (that has a higher noise level than the circular Stokes polarization which provides the $B_{\rm los}$ component). This value was determined by making a series of test calculations of $R$ for HARP/AR full disk passage for threshold values in the range $150-500$\,Mx\,cm$^{-2}$. The chosen value of 300\,Mx\,cm$^{-2}$ corresponds to the threshold making $\log\left[R\left(B_{r}\right)\right]$ closest to $\log\left[R\left(B_{\rm los}\right)\right]$ when the regions are located within 45{\degree} of central meridian.

\subsection{Fourier Spectral Power Index}\label{ss:alpha}

The spectral power index, $\alpha$, (also referred to as exponent) corresponds to the power-law exponent achieved in fitting the function $E\left(k\right) \propto k^{-\alpha}$ to the 1D power spectral density $E\left(k\right)$ in magnetograms. This field-fractality related property parameterizes the power contained in magnetic structures at spatial scales $l = k^{-1}$ belonging to the inertial range of magnetohydrodynamic turbulence. Empirically, ARs with high spectral power index display an overall high productivity of flares \citep[see {\it e.g.}][]{2005ApJ...629.1141A,2015SoPh..290..335G}. 
 Historically, $\alpha$ has been calculated using $B_{\rm los}$ under the assumption that it represents the normal-field component at small distances from disk centre. First, a magnetogram is processed using a FFT to yield $E\left(k_{x}, k_{y}\right)$, the 2D power spectral density (PSD) that is the squared absolute value of the FFT. To convert 2D PSD from wave numbers $k_{x}$ and $k_{y}$ to the 1D PSD $E\left(k\right)$ with isotropic wavenumber $k = \left(k_{x}^{2} + k_{y}^{2}\right)^{0.5}$, it is necessary to integrate $E\left(k_{x}, k_{y}\right)$ over angular direction in Fourier space and multiply by a factor of $2 \pi k$. Finally, a power-law fit is performed as a linear fit to the log-log representation of $E\left(k\right)$ {\it vs} $k$ and the slope ({\it i.e.} $\alpha$) is obtained over the inertial range corresponding to $2-20$\,Mm ({\it i.e.} $k=0.05-0.5$\,Mm$^{-1}$). Power spectral indices are calculated for both $B_{\rm los}$ and $B_{r}$ as described here -- no changes were necessary between the two data formats.

\subsection{Effective Connected Magnetic Field Strength}\label{ss:beff}

The effective connected magnetic field strength, $B_{\rm eff}$, is a morphological proxy that aims to quantify strong MPILs at the photosphere of an AR. It was first introduced by \citet{2007ApJ...661L.109G} and modified by \citet{georgoulis2010,Georgoulis_2013}. $B_{\rm eff}$ is based on the idea that an AR may be represented by a dipole with foot-point separation equal to the magnetogram pixel size and total flux equal to the total connected flux of the AR.  First, patches of both magnetic polarities are identified using the partitioning method of \citet{2005ApJ...629..561B}, producing a set of non-overlapping areas with known outlines, magnetic flux content, and flux-weighted centroid positions. Second, a connectivity matrix is defined for the fluxes $\Phi_{ij}$ committed to opposite-polarity partition pairs $ij$. Each connection is given a length, $L_{ij}$, that is the distance between the flux-weighted centroids of the partition pair. Fluxes in the connectivity matrix are found via a simulated annealing method designed to emphasize MPILs. This is achieved through preferably connecting closest opposite-polarity fluxes by minimizing
\begin{equation}
\sum_{ij}\left( \frac{|\mathbf{\mathit{r_{i}}}-\mathbf{\mathit{r_{j}}}|}{l_{\rm max}} + \frac{|\Phi_{i}+\Phi_{j}|}{|\Phi_{i}|+|\Phi_{j}|}\right) \ ,
\end{equation}
where $r$ and $\Phi$ are the position vector and flux of each partition, $i$ and $j$ indicate positive- and negative-flux partitions, respectively, $l_{\rm max}$ is the diagonal length of the magnetogram, and the sum is performed over all opposite-polarity pairs.

For flux-balanced ARs, a connectivity matrix contains only opposite-polarity connections after the annealing process. However, ARs are rarely flux balanced, mostly due to large-scale connections to flux beyond the FOV. To rectify this, a ring of ``mirror flux'' is placed a large distance from the AR. Connections between excess AR flux and the mirror flux are treated as open and not included in $B_{\rm eff}$. From the connectivity matrix, $B_{\rm eff}$ is measured in magnetic intensity units as
\begin{equation}
B_{\rm eff}=\sum_{i}\sum_{j}\frac{\Phi_{ij}}{L^{2}_{ij}} \ .
\end{equation}
To calculate $B_{\rm eff}$, the entire AR must be taken into account while partitions belonging to the quiet Sun or small isolated partitions (that do not contribute to flare activity) must be excluded. To this end, minimum thresholds are imposed on flux density (100\,Mx\,cm$^{-2}$), partition area (40\,pixels, $\approx$\,5.3\,Mm$^{2}$ from SHARP CEA pixel size of  0.03$\degree$), and enclosed flux ($5\times 10^{19}$\,Mx), with values chosen following AR time-series tests. Threshold selection also affects calculation speed; lower thresholds include larger portions of magnetic flux, producing more partitions at the expense of slower calculations. Since $B_{\rm eff}$ is a quantity calculated from the normal-field component, the same thresholds are used for both $B_{r}$ and $B_{\rm los}$ to compare the values produced, keeping in mind that in principle $B_{r}$ values are higher than the corresponding $B_{\rm los}$ when considering ARs far from disk centre. According to \citet{2008ApJ...688L.107B}, $B_{\rm eff}$ can be seen as an $R$-value with the flux-weighting given by the connectivity matrix. Therefore some correlation between these two properties can be expected.

\subsection{Ising Energy}\label{ss:isinge}

The term Ising energy originates from the solid state physics model with the same name. This model is use to characterize the state of magnetic systems in which opposite polarities can have short and long range interactions, such as ferromagnetic materials. This quantity was proposed as a measure of AR magnetic complexity in MDI LOS magnetograms by \citet{ahmed2010} and corresponds to interaction energy between pairs of magnetic polarities. First, flux values are byte-scaled to $0-255$. Second, low values ($0-30$) represent strongest negative fields and are flagged with $-1$, while high values ($230-255$) represent strongest positive fields and are flagged with $+1$; intermediate values ($31-229$) represent lower absolute field strength and are set to 0 and ignored. Finally, Ising energy is calculated as

\begin{equation}
E_{\rm Ising}=-\sum_{ij}\frac{S_{i}S_{j}}{d^{2}} \ ,
\label{ising}
\end{equation}
where $S_{i}=+1$ and $S_{j}=-1$ for positive and negative pixels, respectively, and $d$ is distance between opposite-polarity pixel pairs $ij$. $E_{\rm Ising}$ correlates with AR flare productivity according to \citet{ahmed2010}, but has not been implemented in the forecasting literature. For the calculation of Ising energy, it is necessary to define ranges that correspond to strong positive and negative pixels. Unlike the byte-scaling applied in \citet{ahmed2010}, an absolute flux threshold is used here; only pixels of absolute value greater than 100\,Mx\,cm$^{-2}$ are considered, and flagged with $+1$ ($-1$) if their value is positive (negative).
 
$E_{\rm Ising}$ and $B_{\rm eff}$ show a similar functional form. However, in the former, unipolar magnetic elements correspond to strong pixels (field strength $>$ 100\,Mx\,cm$^{-2}$) while for latter, the same are represented by non-overlapping partitions. For Ising energy, the  connectivity accounts for all possible connections between strong pixels pairs, without, however, taking into account the magnetic flux of these pixels. Some degree of correlation between $E_{\rm Ising}$ and $B_{\rm eff}$ is expected. Thus, by definition, Ising energy is a non-zero quantity as long as there are at least two strong opposite polarity pixels within a HARP but at least two sizable, non-overlapping opposite polarity magnetic partitions are required for a non-zero $B_{\rm eff}$.

\section{Results}\label{s:results}

A total of $12,773$ SHARP magnetograms were available in the selected sample. This sample was then filtered to leave only those HARP numbers that were associated to (one or more) NOAA-numbered regions at any time during the HARP disk passage. A total of $3,999$ SHARPs ($31.3$\%) are included in the results presented in this section. On occasions, the dynamic evolution of active regions within a HARP associated to multiple NOAA regions can lead to the splitting of the HARP into two or more patches. This particular scenario is important to consider if the full disk passage of HARPs is studied at any cadence. However, the data sample in this study was randomly selected, and therefore the evolution of any particular region is unlikely to be included.

Results from this statistical study are shown with emphasis on four aspects: differences in property values from using $B_{\rm los}$ or $B_{r}$ data (Section~\ref{ss:comp}); variation of properties with AR longitudinal position (Section~\ref{ss:var_pos}); correlations between properties (Section~\ref{ss:par_correl}); flaring association of individual properties (Section~\ref{ss:flaring}).

\subsection{LOS- {\it versus} Radial-field Comparison}\label{ss:comp}

\begin{figure}[!h]
 \centerline{\includegraphics[width=0.5\textwidth,clip=]{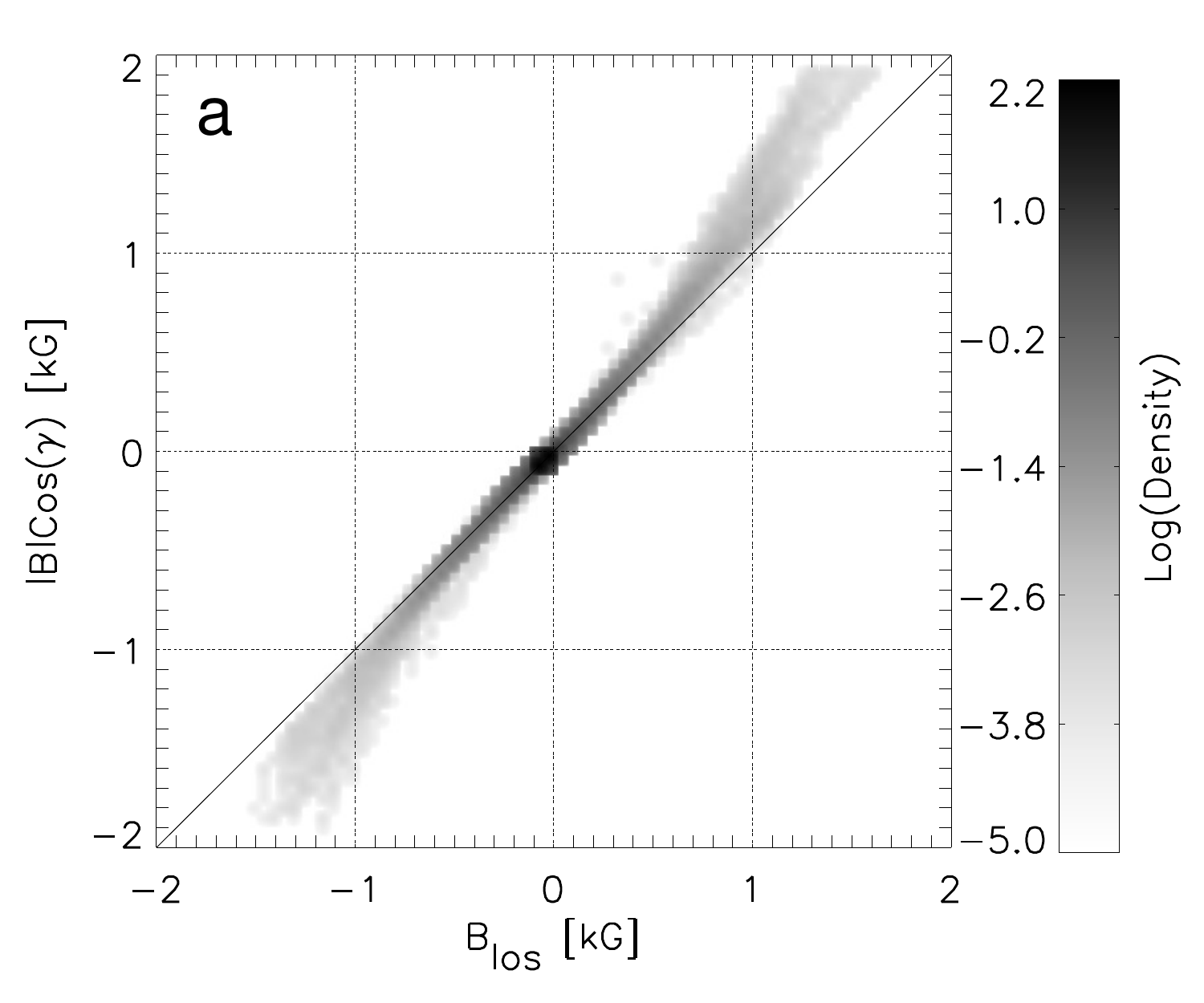}
 \includegraphics[width=0.5\textwidth,clip=]{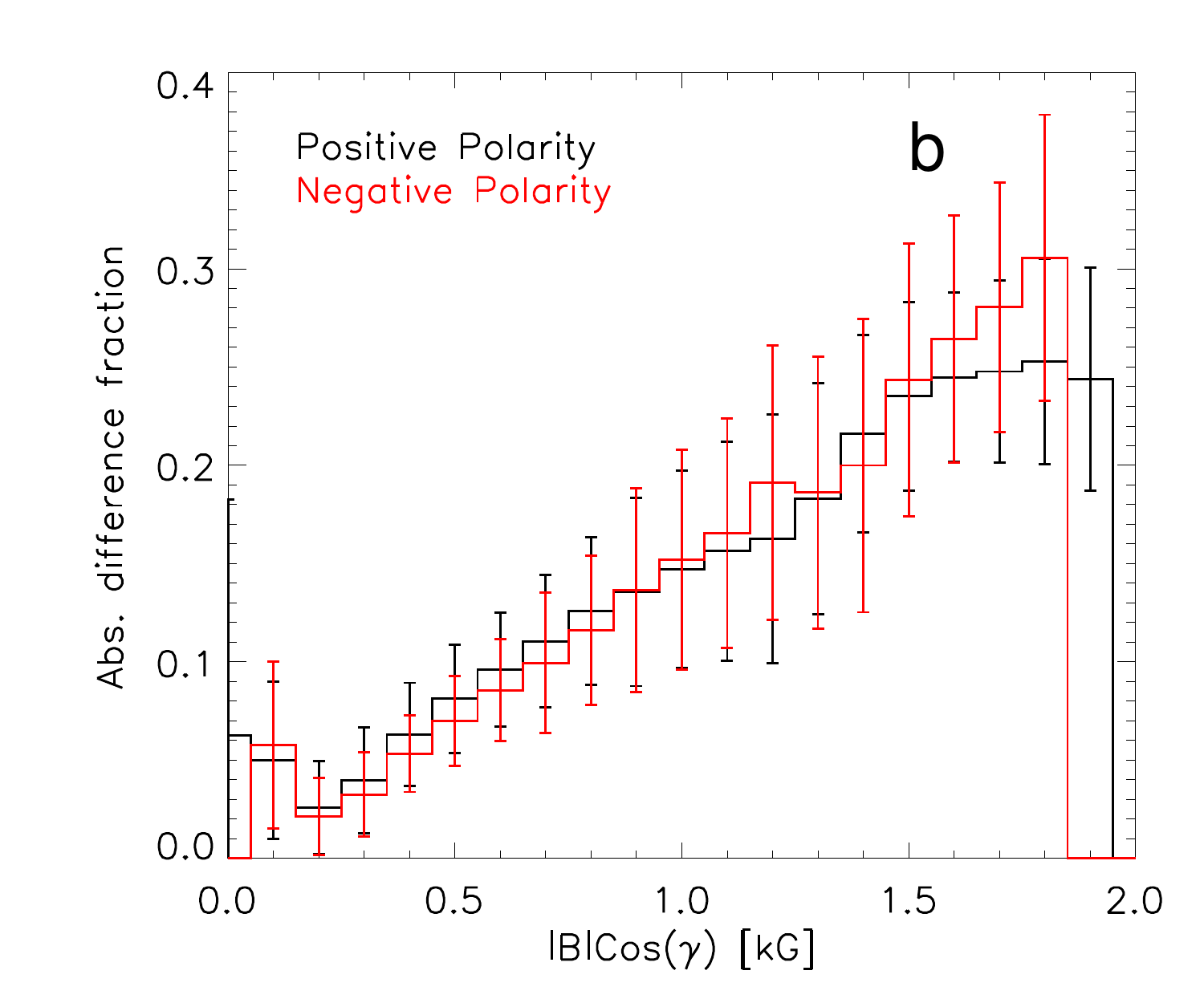}} 
 \caption{Pixel-by-pixel comparative density histogram (a) between the SHARP $B_{\rm los}$ signal and the LOS signal recovered from the Milne-Eddington inversion, $|B|\cos(\gamma)$. Average absolute difference percentage (b) as a function of $|B|\cos(\gamma)$ for each polarity separately. Data used to produced these plots correspond to all HARPs present on 6 August  00:00\,UT. $B_{\rm los}$ seems to systemically underestimate the field in pixels with absolute strength greater than 500 G. The maximum observed difference between the two signals for strong-field pixels ($\approx$ 1.8 kG) is 25\% and 30\% for positive and negative polarity pixels, correspondingly. Error bars correspond to the 1$\sigma$ values.}
 \label{fig:blos_comp}
\end{figure}

Most properties studied here are strictly speaking defined in terms of the surface-normal ($B_{\rm n}$) component of the photospheric field. In their original implementations, properties are calculated either using only $B_{\rm los}$ (if restricted to disk centre) or $\mu$-angle corrected $B_{\rm los}$. The availability of $B_{r}$ (equivalent to $B_{\rm n}$) and $B_{\rm los}$ in the SHARP data products allows direct comparison between the resulting properties that each produces. However, SHARP $B_{\rm los}$ and $B_{r}$ magnetograms are derived from different polarization signal ({\it i.e.} different detectors) and therefore the LOS component from the ME inversion ($|B|\cos(\gamma)$; $|B|$ = field strength, $\gamma$ = inclination angle respect to the LOS) does not necessarily  match $B_{\rm los}$. Therefore, comparison between $|B|\cos(\gamma)$ and $B_{\rm los}$ in the plane-of-sky coordinate system should be done before proceeding with the AR-property calculations. Figure~\ref{fig:blos_comp}{a} presents the density histogram from a pixel-by-pixel comparison of all HARP regions on 6 August 2014 at 00:00\,UT. This date was chosen because multiple regions of different sizes are present and more or less distribute all over the solar disk. Figure~\ref{fig:blos_comp}{b}, on the other hand, shows the average percentage difference between the two signals as a function of $|B|\cos(\gamma)$, for each polarity separately. Figure~\ref{fig:blos_comp}{a} shows that for pixels with $>$ 500 G, $B_{\rm los}$ underestimates the field strength. Positive and negative polarities (black and red curves in Figure~\ref{fig:blos_comp}{b}, respectively) show similar average difference between $B_{\rm los}$ and $|B|\cos(\gamma)$ values, up to $\approx$ 1.5 kG, when negative polarities are observed to have larger difference. At the highest common value, positive and negative polarities are, on average, underestimated by $B_{\rm los}$ by 25\% and 30\%, respectively. Although, in Figure~\ref{fig:blos_comp}{a} it is observed that such big differences display a density of only about 10$^{-5}$.

\begin{figure}[!h]
 \centerline{\includegraphics[width=0.5\textwidth,clip=]{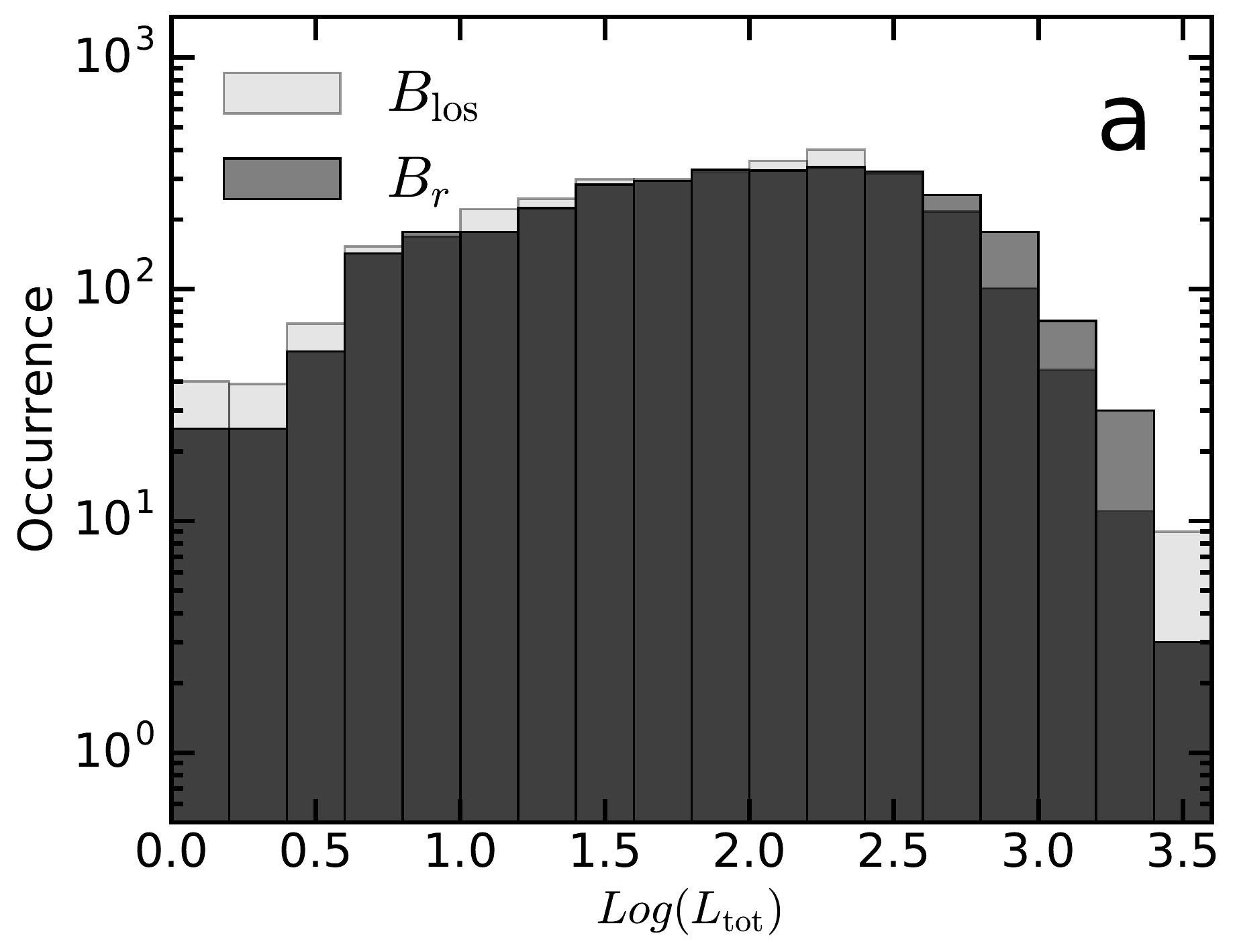}\includegraphics[width=0.5\textwidth,clip=]{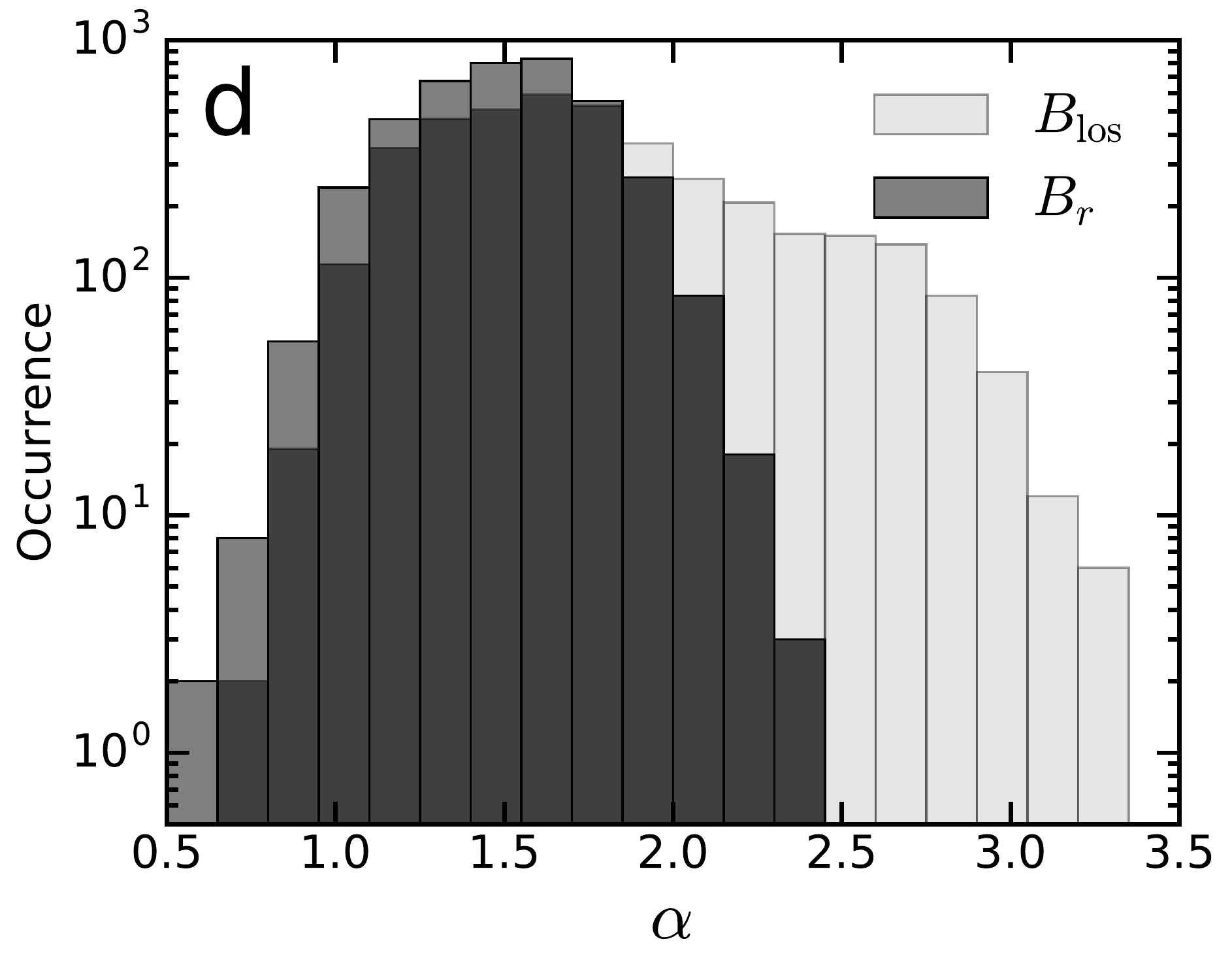}} 
 \centerline{\includegraphics[width=0.5\textwidth,clip=]{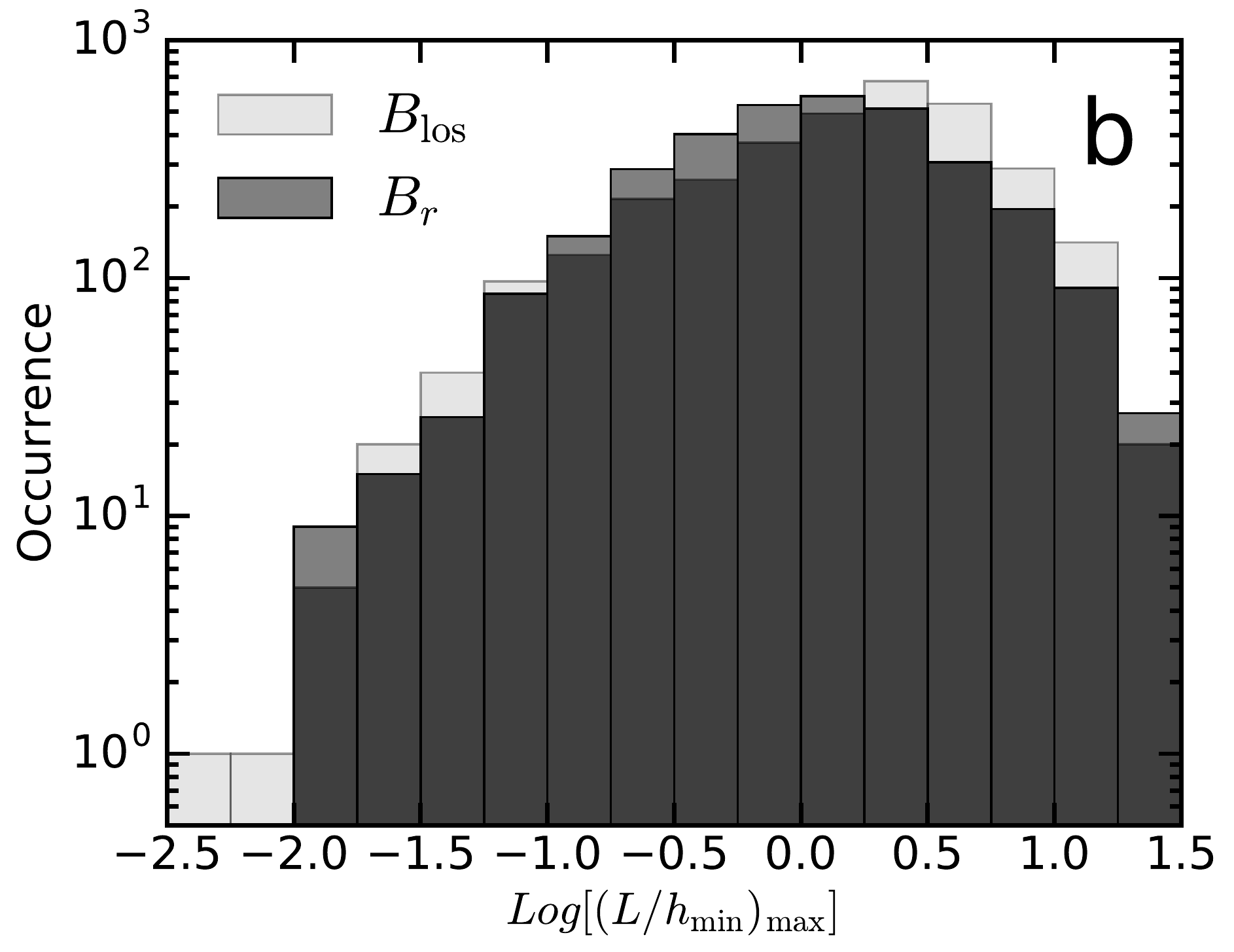}\includegraphics[width=0.5\textwidth,clip=]{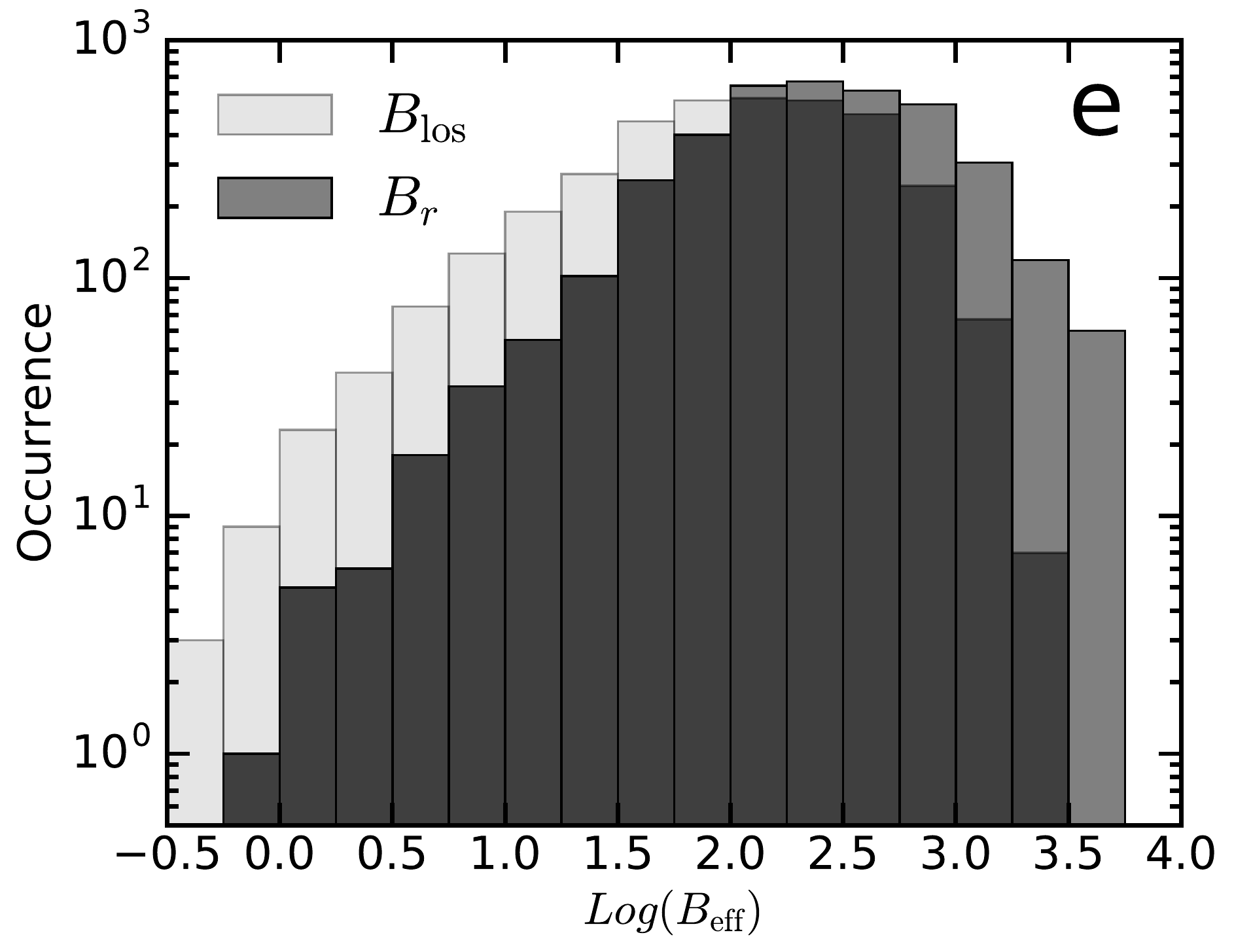}}
 \centerline{\includegraphics[width=0.5\textwidth,clip=]{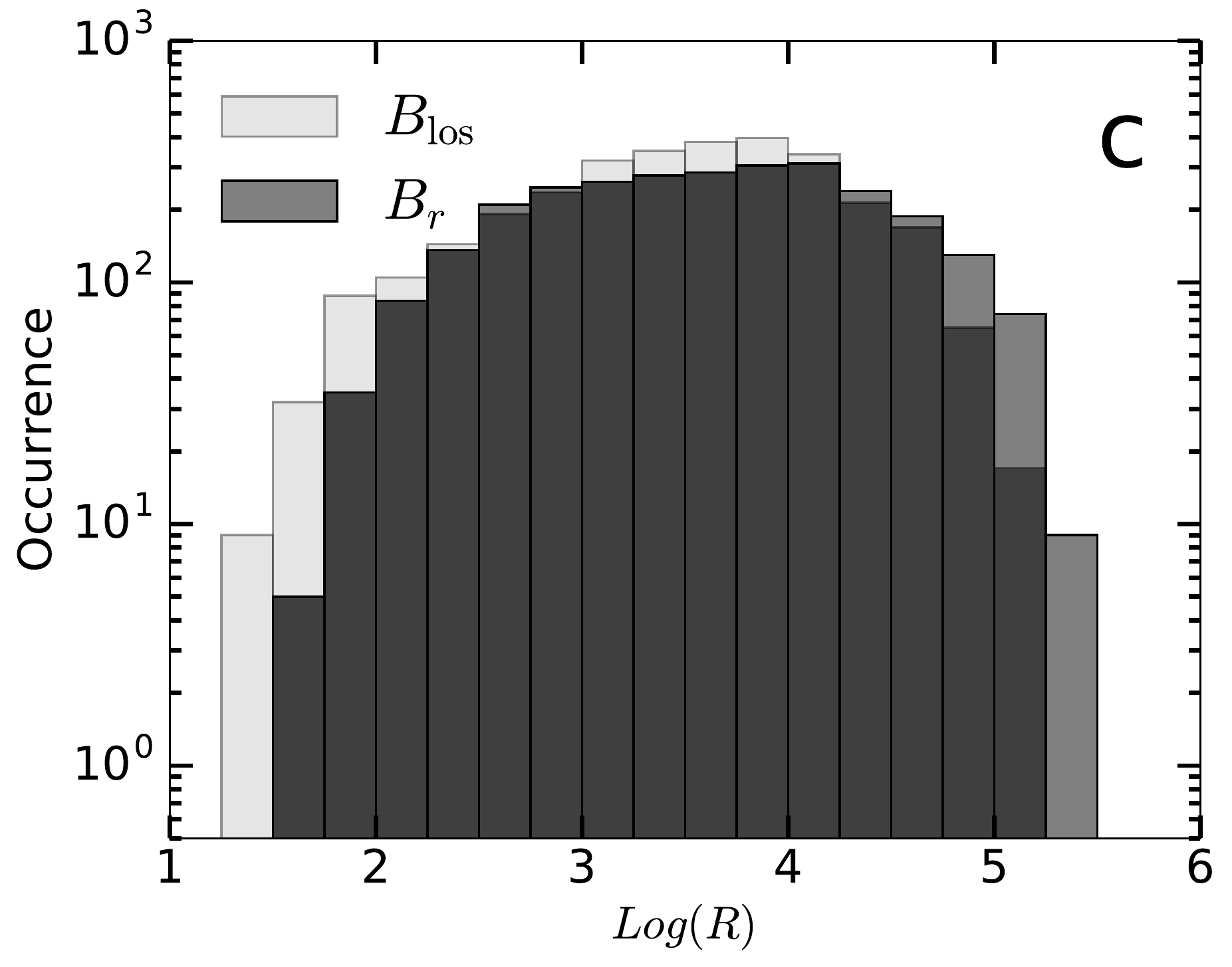}\includegraphics[width=0.5\textwidth,clip=]{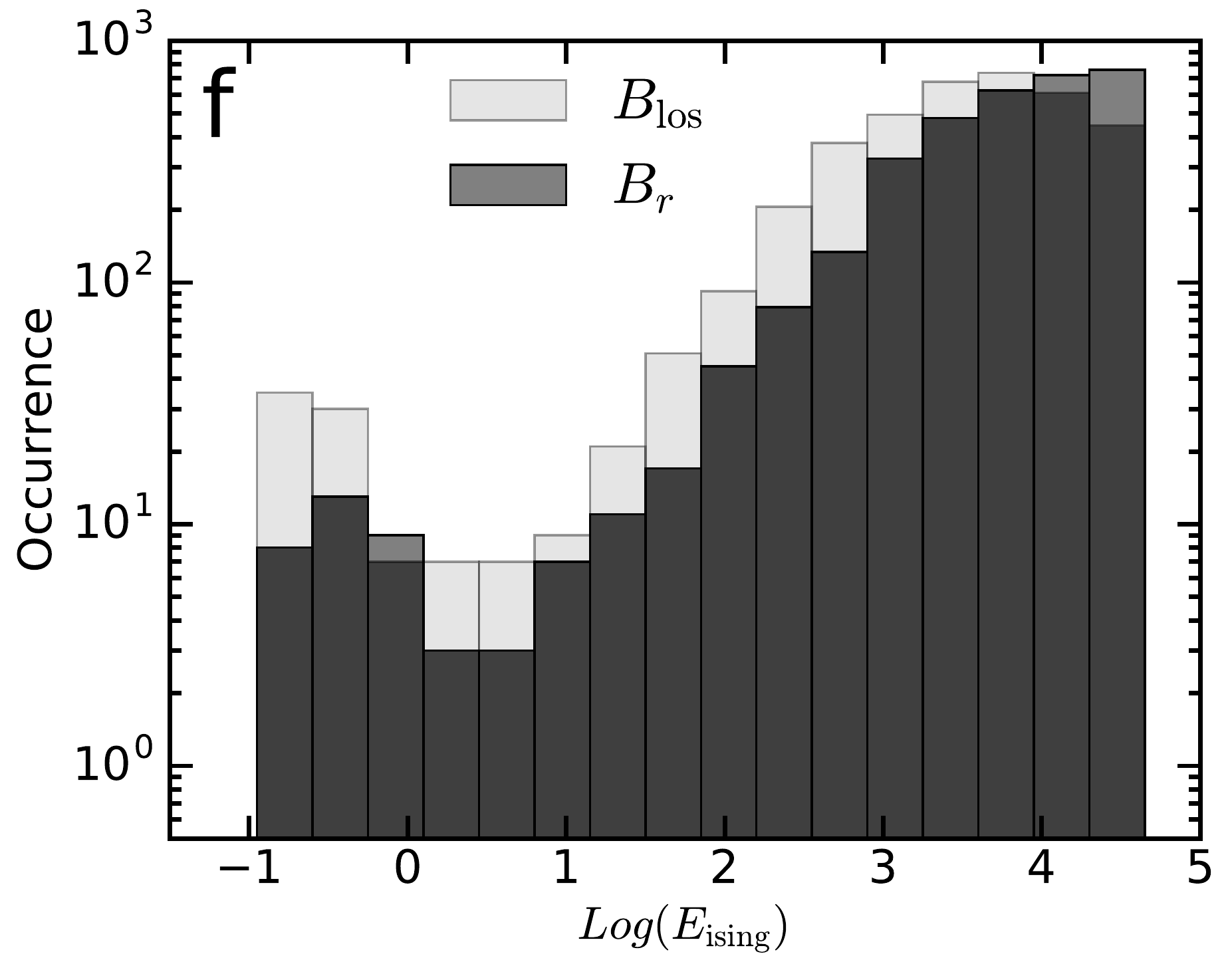}}
 \caption{Frequency distributions of AR properties: (a) total $^{\star}$MPIL length, $L_{\rm tot}$; (b) maximum ratio of $^{\star}$MPIL length to minimum height of critical decay index,  $\left(L/h_{\rm min}\right)_{\rm max}$; (c) Schrijver's $R$ value; (d) Fourier spectral power index, $\alpha$; (e) effective connected field strength, $B_{\rm eff}$; (f) Ising energy, $E_{\rm Ising}$. In all panels, grey-scale shading displays the distributions from $B_{\rm los}$ (light grey) and from $B_{r}$ (mid-grey), while regions of distribution overlap are indicated in dark grey.}
 \label{fig:histograms}
\end{figure}

Thus, the relatively small difference between the two LOS signals allows the use of the readily available SHARP $B_{\rm los}$ in this study. In addition, the well-known lower noise level (5--10 G; compared to 60--150 G of the ME inversion LOS) and the difficulty to recover the LOS component from the CEA fields, make the SHARP $B_{\rm los}$ more useful in real time forecasting applications. Figure~\ref{fig:histograms} contains six panels displaying the distribution of values for all properties calculated. In all panels of Figure~\ref{fig:histograms}, grey-scale shading displays the distributions from $B_{\rm los}$ (light grey) and from $B_{r}$ (mid-grey), while regions of distribution overlap are indicated in dark grey.

Figure ~\ref{fig:histograms} (panels a-c) presents the histograms for the MPIL-related properties. Distributions of the total length of strong-gradient MPILs ($L_{\rm tot}$; Figure~\ref{fig:histograms}{a}) show that very similar behaviour is found from $B_{\rm los}$ and $B_{r}$ -- both distributions cover a similar range of values. Lower occurrence at high values of $L_{\rm tot}$ reflects the drop-off in observation of large/complex ARs while the decrease in occurrence for small values appears to be related to the fact that strong MPILs are rarely found in only one or two magnetogram's pixels. For $\left(L/h_{\rm min}\right)_{\rm max}$, the maximum ratio of MPIL length to minimal height of critical decay index (Figure~\ref{fig:histograms}{b}), as with $L_{\rm tot}$, both $B_{\rm los}$ and $B_{r}$ show similar shape and span similar ranges of values. Both distributions for this property are slightly skewed to higher values. This skewness implies that most probable value of this property comes from ARs that display $L > h_{\rm min}$. This property combines two AR measures ($^{\star}$MPIL length and minimum height of critical decay index), so it is difficult to assess the variation between $B_{\rm los}$ and $B_{r}$ as both properties vary between data types. In the distributions of Schrijver's $R$ (Figure~\ref{fig:histograms}{c}) the main observed  difference is the $B_{r}$ distribution being slightly shifted towards higher values in comparison to the $B_{\rm los}$ distribution. This difference results from the higher flux threshold used with $B_{r}$, which removes some of the low R-values (log($R$) = 1 -- 2.5), and from the systematic underestimation of field strength by $B_{\rm los}$ (Figure \ref{fig:blos_comp}), which affects those regions producing larger values (log($R$) $>$ 3).

In Figure~\ref{fig:histograms}{d}, the $\alpha$ (Fourier power spectral index) distributions are shown. Notably, the range of values from $B_{\rm los}$ data is reduced when using $B_{r}$ data. \citet{2015SoPh..290..335G} reported values of $\alpha>2.5$ from $B_{\rm los}$ data as AR NOAA 11158 approached the limb, implying that the values of $\alpha$ which are most different when using $B_{r}$ likely correspond to ARs far from disk centre. This can be verified in Section \ref{ss:var_pos} since the property dependence on AR location will be analyzed. The modification of such values appears to cause a slight shift to lower values in the $B_{r}$ distribution, although both distributions display a relatively well-defined peak in the bin 1.55 -- 1.70. This property-value range includes $\alpha$ = 1.67 $\approx$ 5/3, the Kolmogorov exponent \citep{1941DoSSR..30..301K}, implying that a large portion of studied HARPs correspond to ARs which photospheric plasma is in (or near to) a fully developed hydrodynamical turbulence.

Figures~\ref{fig:histograms}{e-f} display the histograms of the magnetic connectivity properties, respectively. Distributions of the effective connected magnetic field strength, $B_{\rm eff}$, display a similar behaviour as Schrijver's $R$ value in Figure~\ref{fig:histograms}{c} ({\it i.e.} a slight shift to larger values when using $B_{r}$ data). This behaviour is in correspondence with the field-strength underestimation of $B_{\rm los}$, and puts in evidence the correlation between $B_{\rm eff}$ and $R$. For the Ising energy, $E_{\rm Ising}$, distributions from $B_{\rm los}$ and $B_{r}$ span the same range and show a similar distinctive shape, unlike any other properties. First, there is a clear tendency of the occurrence to increase with the $E_{\rm ising}$ value -- most NOAA-numbered HARPs show complex connectivity dominated by strong ({\it i.e.} $>$~100 Mx cm$^{2}$) opposite-polarity pixels pairs close to each other. On the other hand, the presence of a peak at low values of 0.1--1.0 pixel$^{-2}$ is due to the definition of $E_{\rm ising}$ -- unlike $B_{\rm eff}$, only the distance separating pixels is used and not magnetic flux. Therefore, some of the HARPs containing early stages of AR formation ({\it i.e.} small numbers of significant-flux pixels that separate during spot formation, before additional flux emergence partially fills the AR interior) can produce the number of low-value regions observed in that peak.

\subsection{Variation with AR Longitudinal Position}\label{ss:var_pos}

\begin{figure}
 \centerline{\includegraphics[width=0.55\textwidth,clip=]{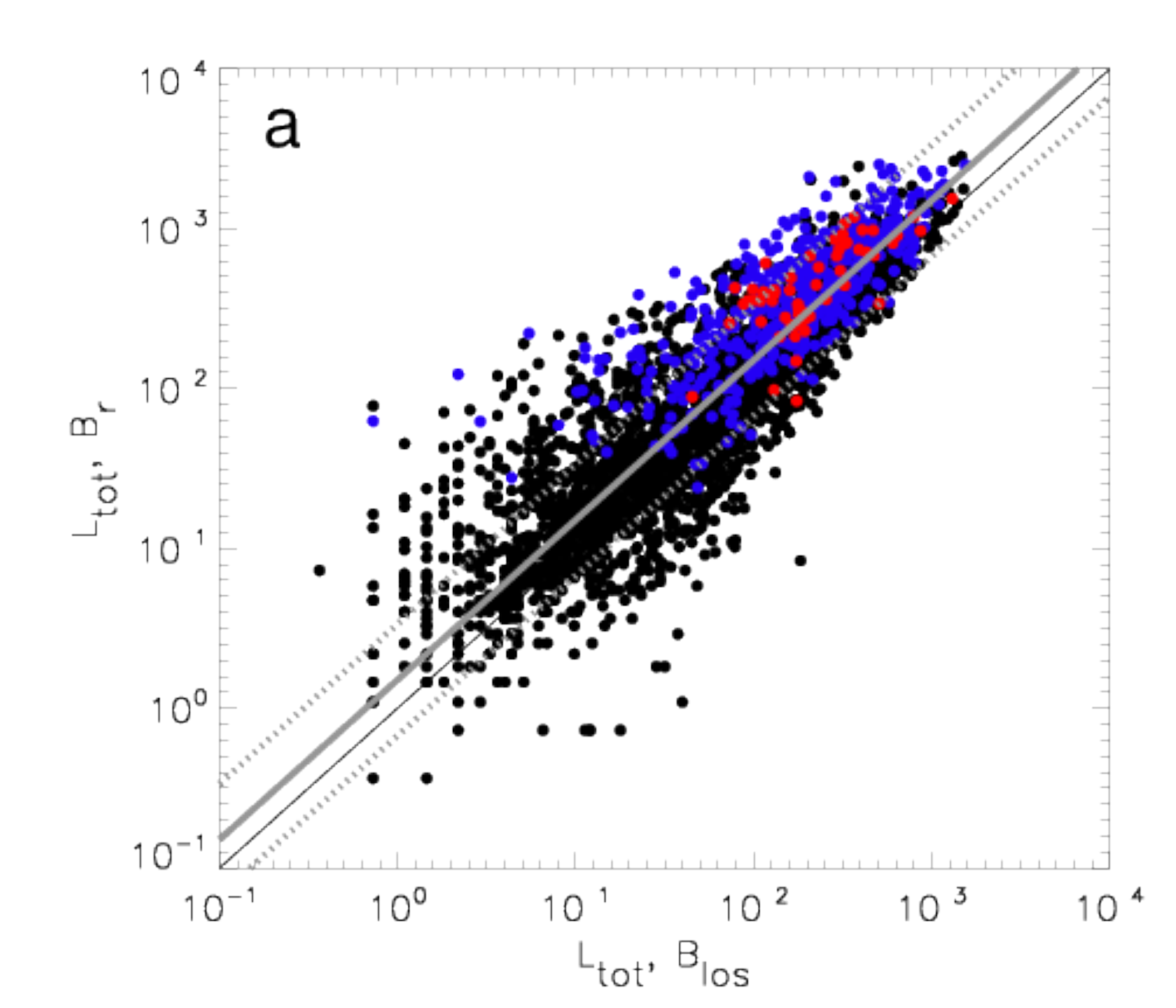}\includegraphics[width=0.55\textwidth,clip=]{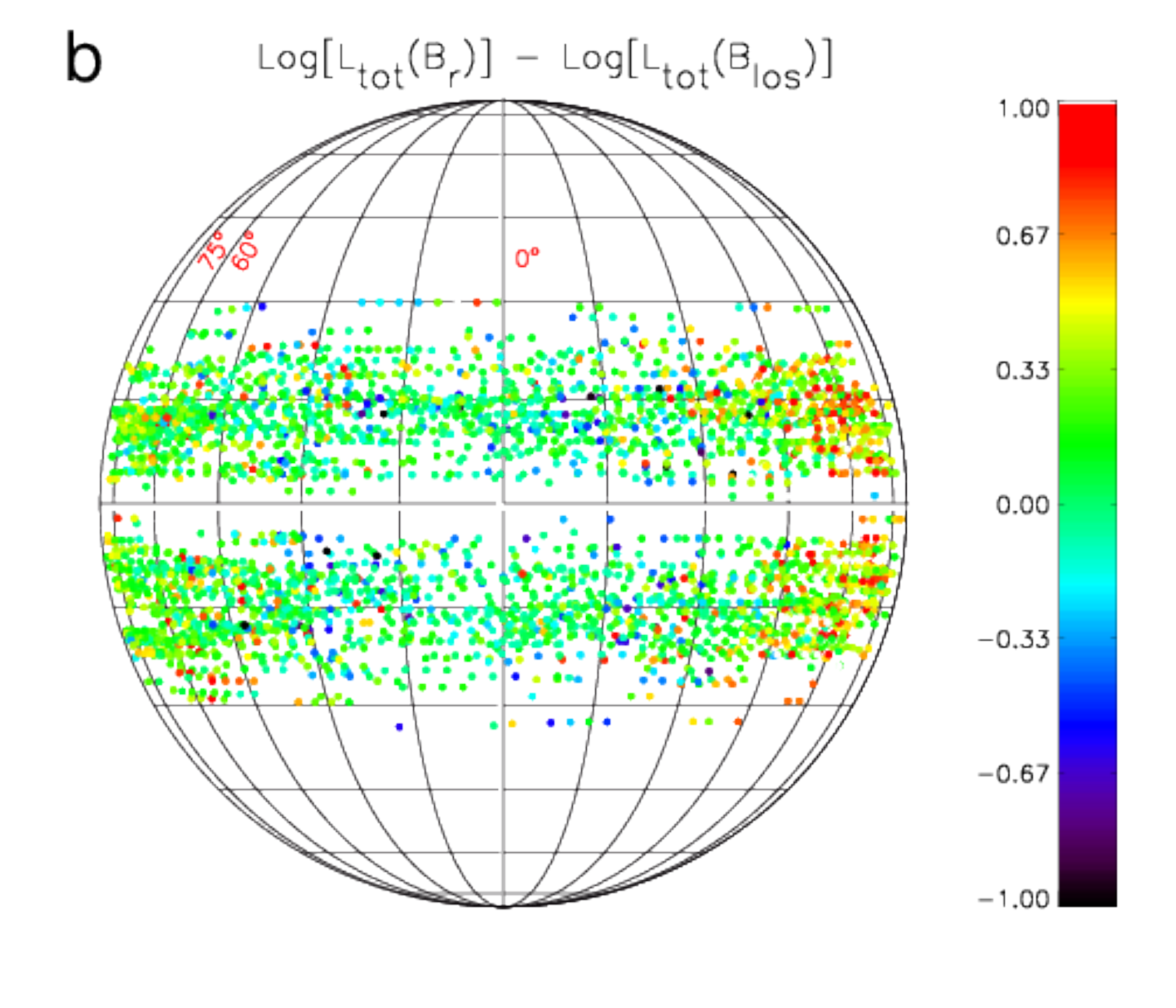}}
 \centerline{\includegraphics[width=0.55\textwidth,clip=]{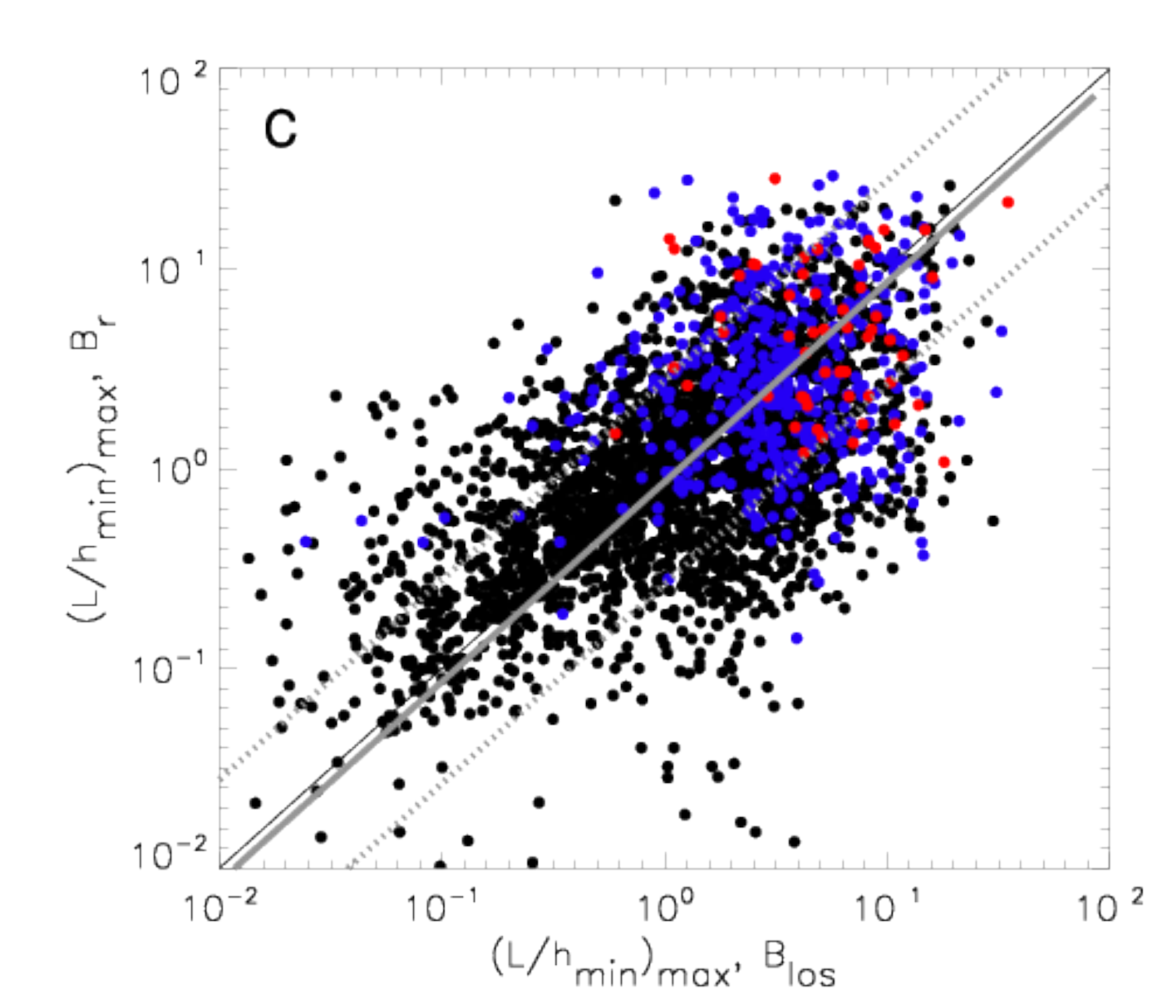}\includegraphics[width=0.55\textwidth,clip=]{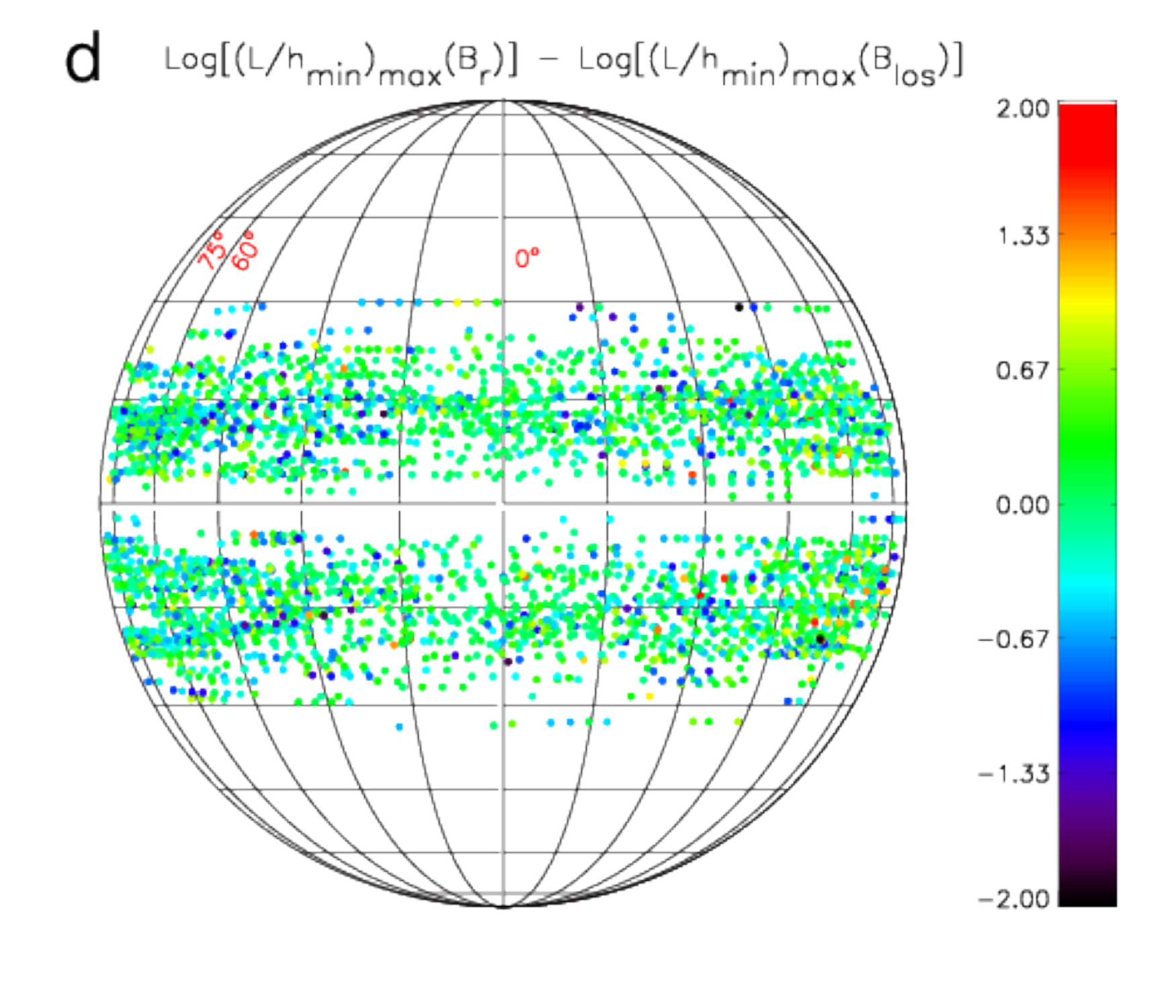}}
  \centerline{\includegraphics[width=0.55\textwidth,clip=]{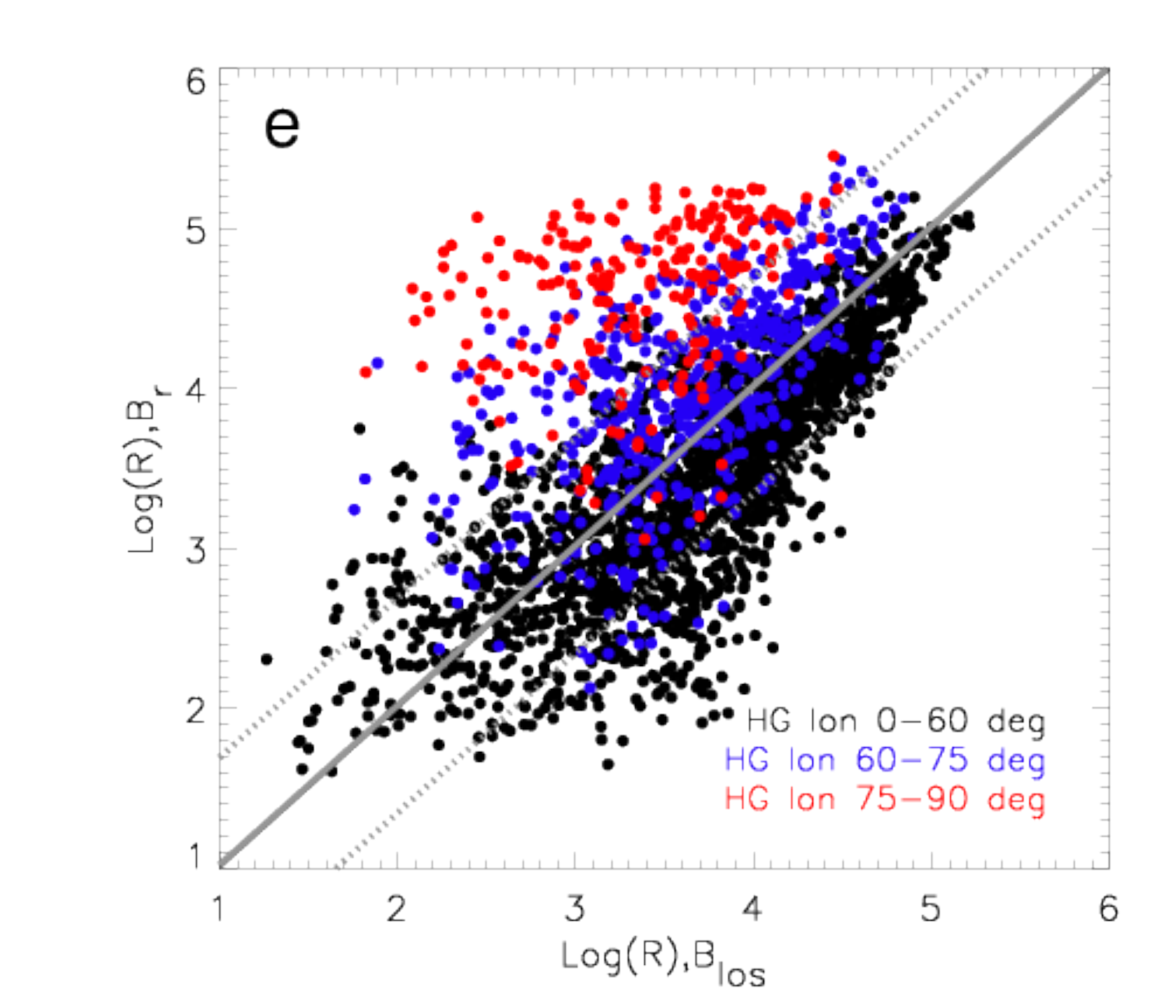}\includegraphics[width=0.55\textwidth,clip=]{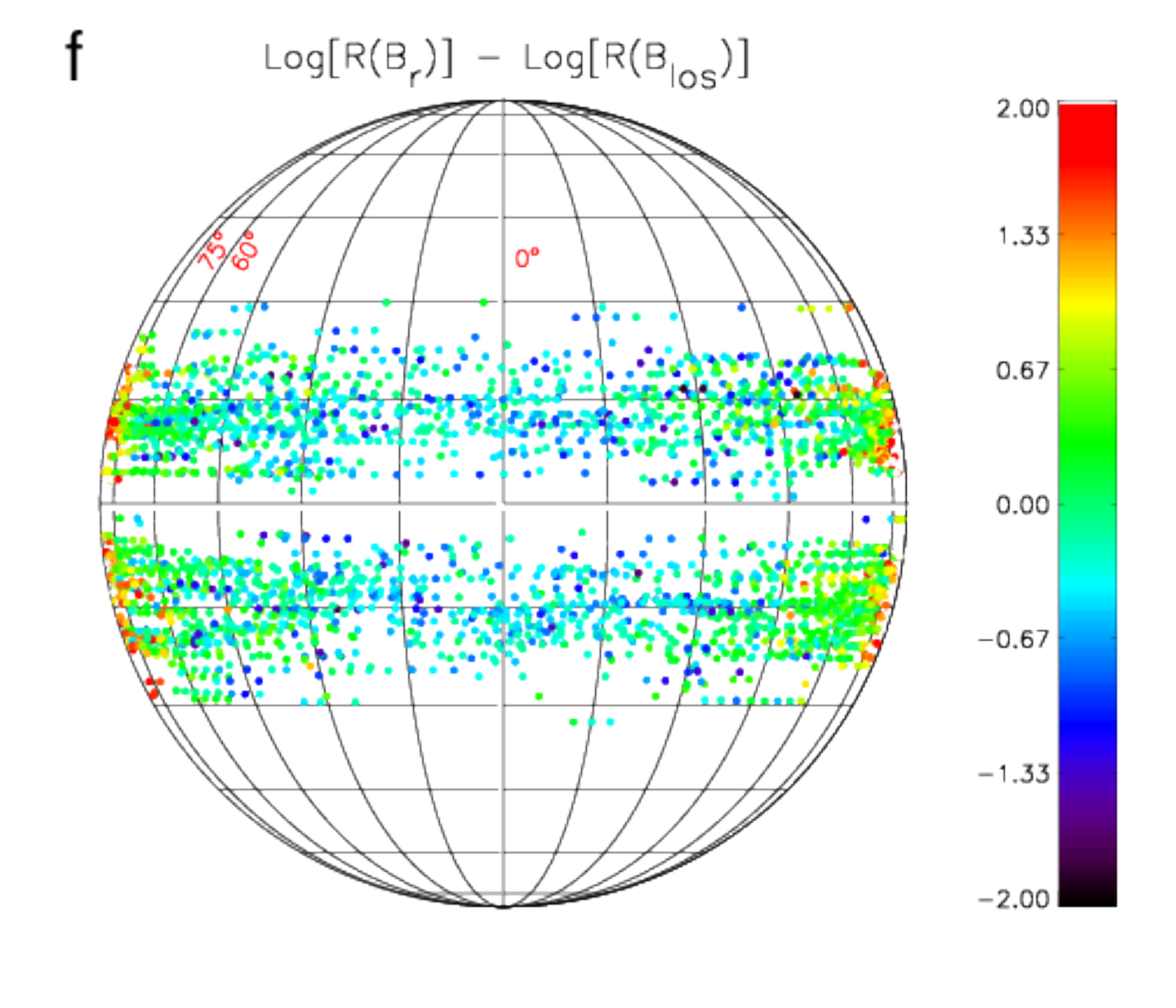}}
 \caption{Differences between $B_{\rm los}$- and $B_{r}$-derived properties: (a)--(b) $L_{\rm tot}$, (c)--(d) $\left(L/h_{\rm min}\right)_{\rm max}$; (e)--(f) $\log\left(R\right)$. Left-hand panels (a, c, and e) show scatter plots of properties derived from $B_{\rm los}$ \textit{vs} $B_{r}$, where the diagonal line is unity and data points are colour-coded to represent three longitude groups:  $|\phi| < 60\degree$ (black); $60\degree \leqslant |\phi| < 75\degree$ (blue); $|\phi| \geqslant 75\degree$ (red). Right-hand panels (b, d, and f) present locations and magnitude differences between $B_{r}$- and $B_{\rm los}$-derived properties, with each data point at the SHARP centroid position at the observation time and colour representing magnitude of difference between property values from the field components.}
 \label{fig:pos_plot_mpil_di_r}
 \end{figure}

\begin{figure}
 \centerline{\includegraphics[width=0.55\textwidth,clip=]{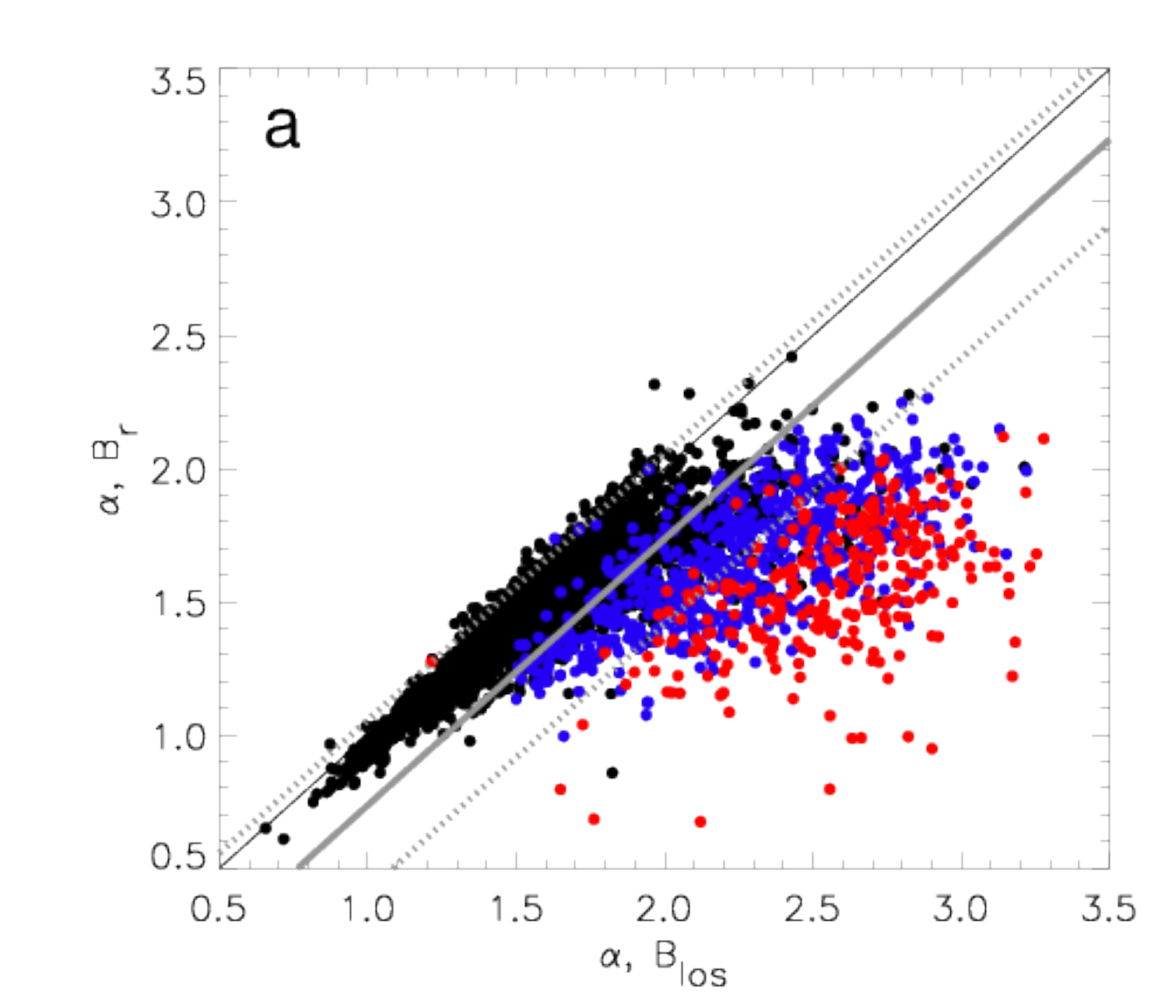}\includegraphics[width=0.55\textwidth,clip=]{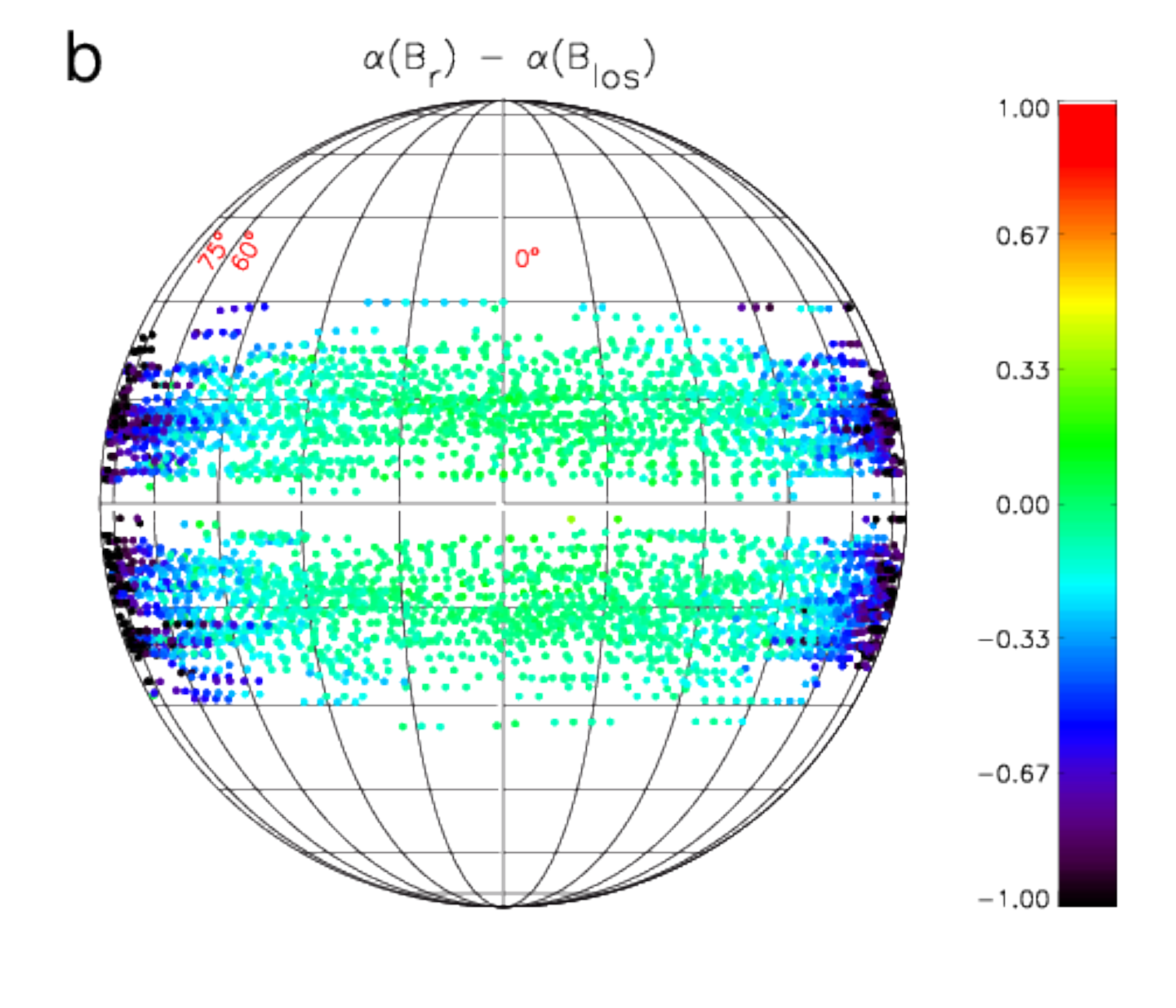}}
 \centerline{\includegraphics[width=0.55\textwidth,clip=]{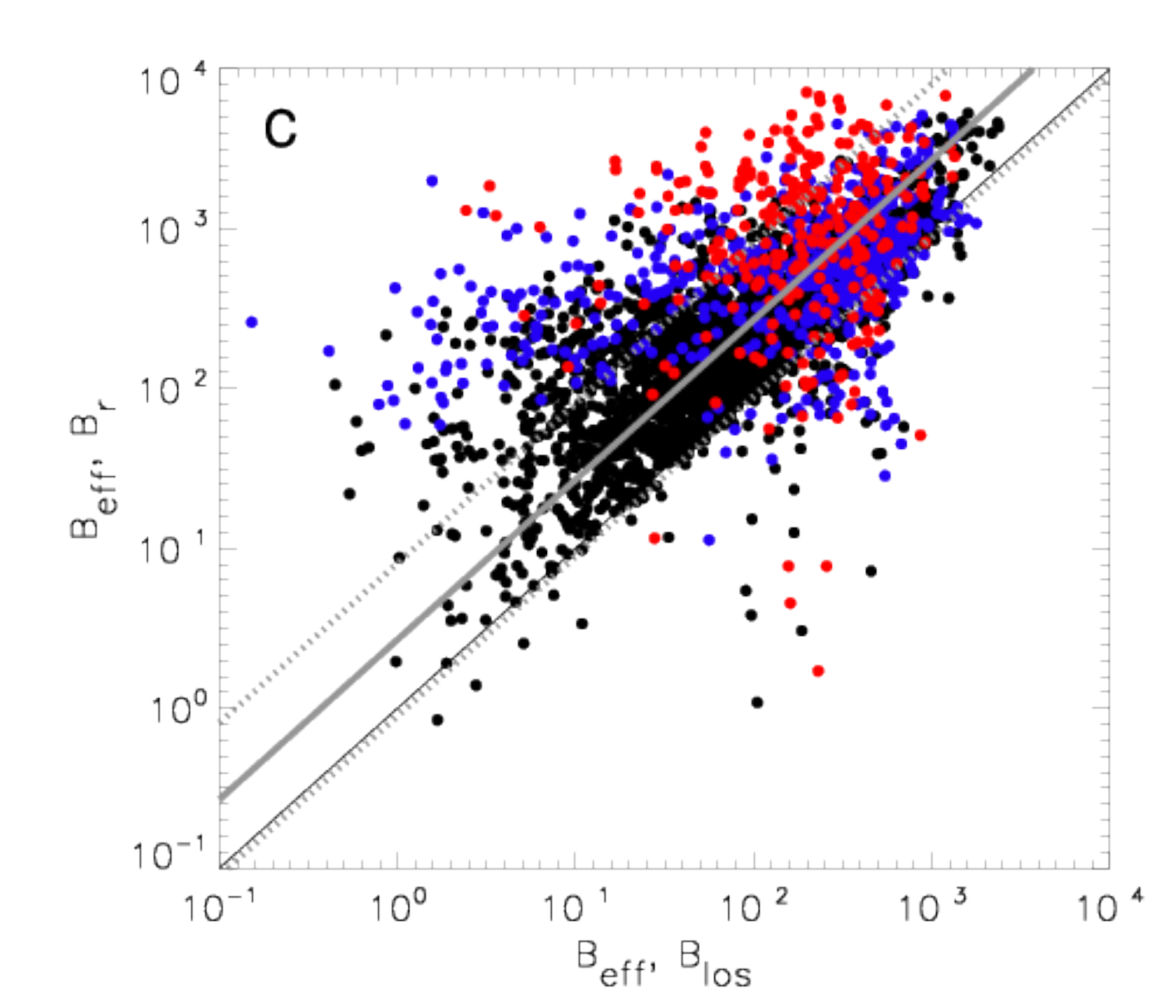}\includegraphics[width=0.55\textwidth,clip=]{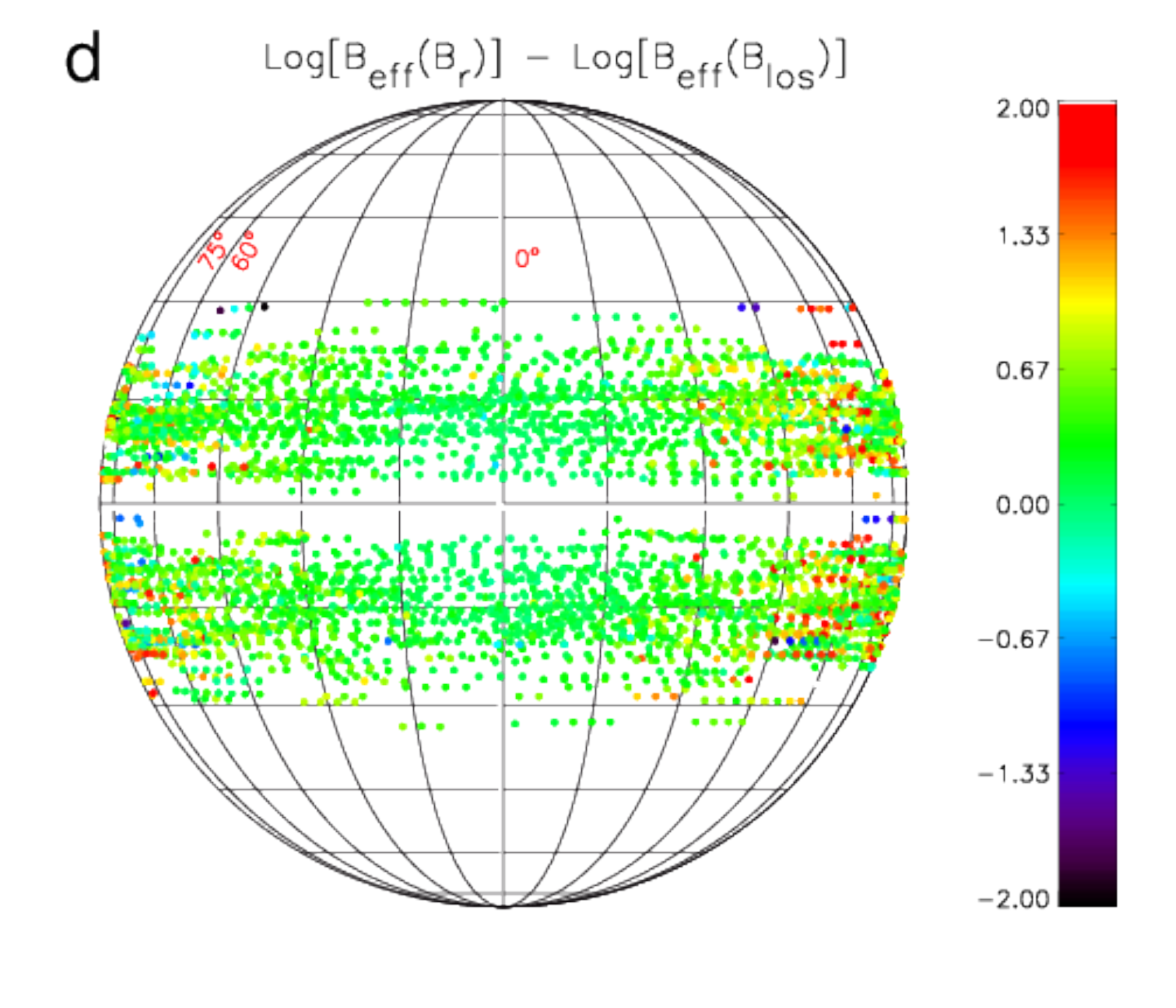}}
  \centerline{\includegraphics[width=0.55\textwidth,clip=]{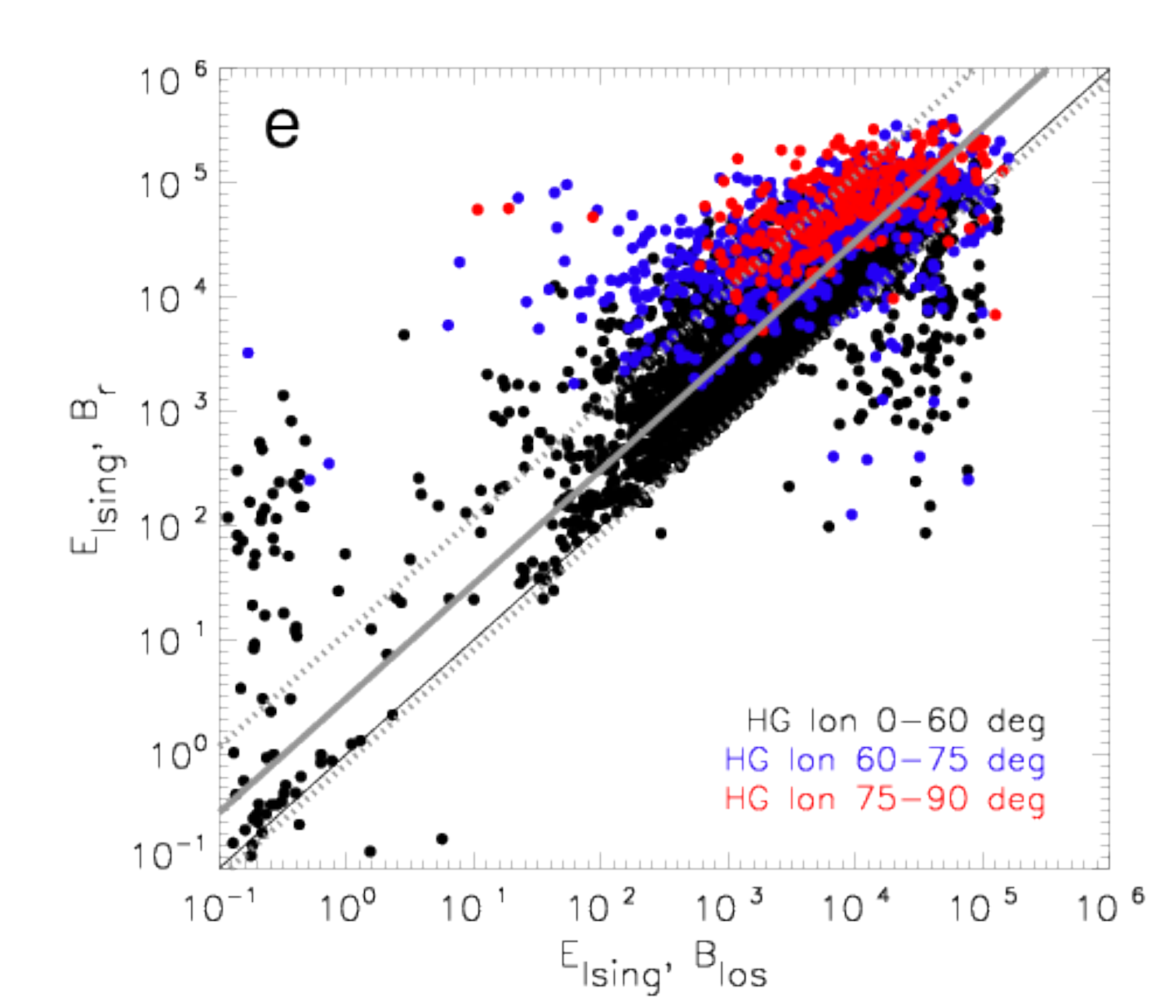}\includegraphics[width=0.55\textwidth,clip=]{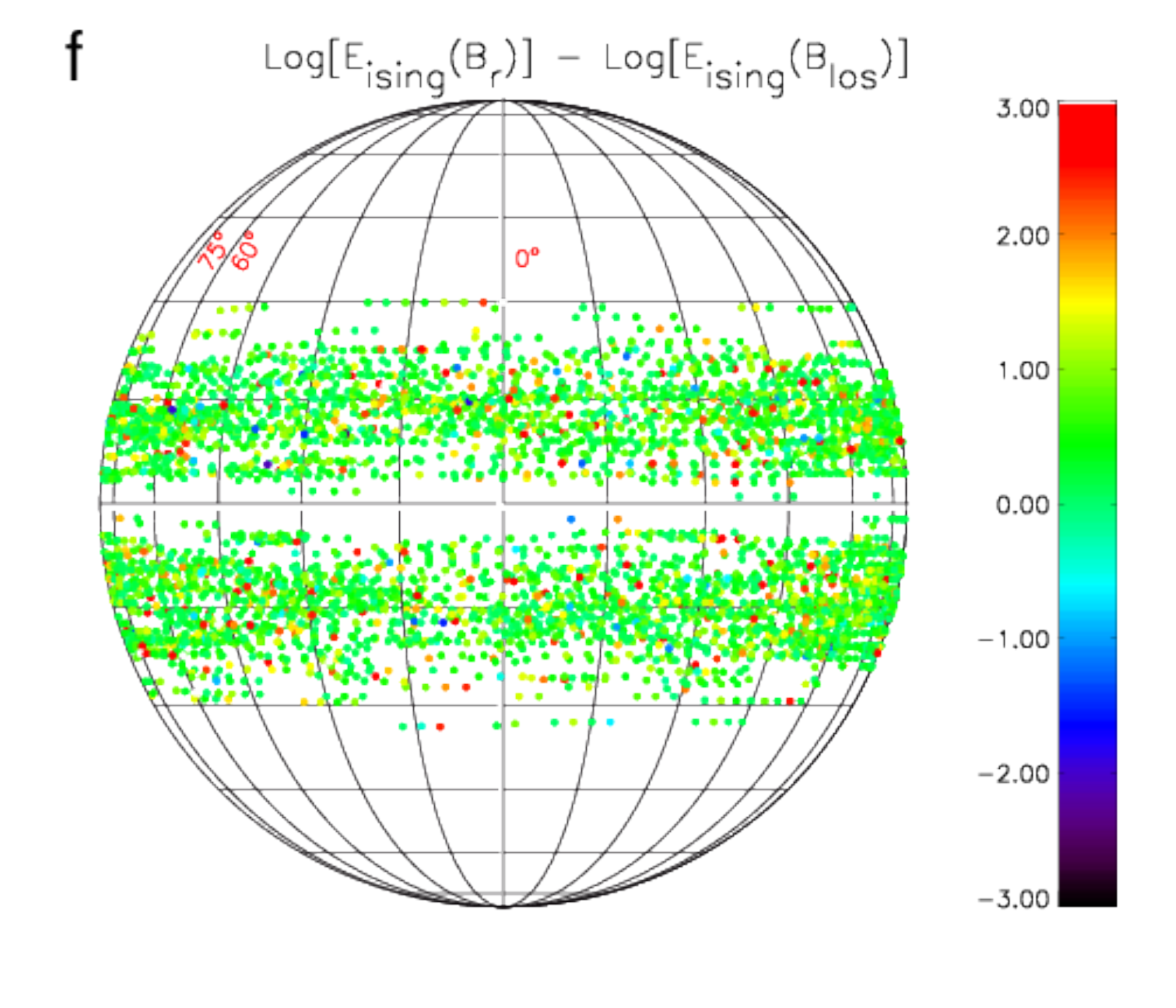}}
 \caption{As Figure~\ref{fig:pos_plot_mpil_di_r}, but for properties: (a)--(b) $\alpha$, (c)--(d) $B_{\rm eff}$; (e)--(f) $E_{\rm Ising}$.}
 \label{fig:pos_plot_alpha_beff_eising}
 \end{figure}

One objective of this study is to investigate the dependence of derived properties on SHARP Heliographic (HG) position. Since AR evolution shows very little variation in latitudinal position, more emphasis is given to the variation of properties with the HG longitude, $\phi$. Figures~\ref{fig:pos_plot_mpil_di_r} and \ref{fig:pos_plot_alpha_beff_eising} present differences between $B_{\rm los}$- and $B_{r}$-derived properties in two forms. First, panels a, c, and e show scatter plots of property values derived from $B_{\rm los}$ \textit{versus} $B_{r}$, where the diagonal black line is unity and data points are colour-coded to represent three groups of SHARP longitudes:  $|\phi| < 60\degree$ (black); $60\degree \leqslant |\phi| < 75\degree$ (blue); $|\phi| \geqslant 75\degree$ (red). Second, panels b, d, and f present locations and magnitudes of difference in $B_{\rm los}$- and $B_{r}$-derived properties, with each data point at the SHARP centroid position at the observation time and colour representing magnitude of difference between property values from the field components. The difference is defined as {\bf $X\left(B_{r}\right) - X\left(B_{\rm los}\right)$} where $X$ can be any of the properties studied, such that when $B_{r}$ yields a larger property value than $B_{\rm los}$ the difference is positive. In all cases except $\alpha$, differences are calculated after taking the logarithm of the property, in order to better visualize differences. In addition, Table \ref{table1} presents maximum and minimum values and the first four moments (mean, standard deviation, kurtosis, and skewness) for the distributions of differences (all $\phi$ groups included). Mean and $\pm \sigma$ values are depicted in Figures~\ref{fig:pos_plot_mpil_di_r} and \ref{fig:pos_plot_alpha_beff_eising} (a, c, e) as solid and dotted grey diagonal lines, correspondingly.

\begin{table}
\caption{Distribution moments of the difference between property values from the field components $B_{\rm los}$ and $B_{r}$.}
\begin{tabular}{llcccccc}%
\hline
Property  & Min & Max & Mean & Std. Dev. & Kurtosis & Skewness \\
\hline
$\log(L_{\rm tot})$ & -1.560 & 2.093 & 0.178 & 0.349 & 2.212 & 0.464 \\
$\log\left[\left(\frac{L}{h_{\rm min}}\right)_{\rm max}\right]$ & -2.451 & 2.173 & -0.067 & 0.509 & 0.768 & -0.130 \\
$\log(R)$ & -1.888 & 2.624 & 0.018 & 0.673 & 0.912 & 0.706 \\
$\alpha$ & -1.953 & 0.350 & -0.264 & 0.321 & 2.650 & -1.698 \\
$\log(B_{\rm eff})$ & -2.129 & 3.235 & 0.430 & 0.478 & 3.552 & 0.958 \\
$\log(E_{\rm Ising})$ & -2.616 & 4.289 & 0.489 & 0.575 & 5.077 & 0.820 \\
\hline
\end{tabular}
\label{table1}
\end{table}

The $L_{\rm tot}$ scatter plot in Figure \ref{fig:pos_plot_mpil_di_r}{a} displays a linear log-log correlation between values calculated from $B_{\rm los}$ and those from $B_{r}$. The distribution of points around the unity diagonal shows a small positive mean, implying that $B_{r}$ values are larger on average. This correlation shows a low level of dispersion, with 2/3 of points (2$\sigma$ -- the distance between dotted grey lines) spanning $\approx$ 0.7, less than an order of magnitude. There appears to be no systematic dependence on longitude as blue and red symbols ($60\degree \leqslant |\phi| < 75\degree$ and $|\phi| \geqslant 75\degree$, respectively) appear to fall within the scatter of black symbols ($|\phi| < 60\degree$). In Figure \ref{fig:pos_plot_mpil_di_r}{b}, the majority of points have differences close to zero (green symbols) and few non-zero values are found for $|\phi| < 30\degree$. At greater $\phi$, $L_{\rm tot}$ values from $B_{r}$ are marginally larger than those from $B_{\rm los}$ (yellow symbols). Overall the largest differences are on the order of tens of Mm, with these magnitudes restricted to $B_{r}$ resulting in larger values than $B_{\rm los}$ (red symbols). The $^{\star}$MPIL detection process is identical for both field components, so larger $B_{r}$ values are due to more pixels passing the $^{\star}$MPIL flux threshold, as a consequence of field-strength underestimation of $B_{\rm los}$. In addition, two characteristics are clearly identified in Figure \ref{fig:pos_plot_mpil_di_r}{b}: i) a systematically larger $B_{r}$ values (positive differences) with increasing $|\phi|$, and ii) asymmetric distribution of values over the disk, with higher differences achieved for western longitudes. These two observations seem to be consistent with the spatial and temporal variations of noise levels in the HMI inverted magnetic field data \citep{2014SoPh..289.3483H}. The noise in the vector field strength exhibits a center-to-limb variation ($\approx$ 60 -- 150 G) that also varies with the satellite orbital velocity. This variation produces an east-west asymmetry in the noise pattern which appears periodically around 1:00, 7:00, 13:00, and 19:00 UT \citep[see Figures 6 and 7 in ][]{2014SoPh..289.3483H}. These times are close to the daily sampling times used in this study. Furthermore, the $^{\star}$MPIL detection algorithm uses a flux threshold of  100 -- 120 Mx cm$^{-2}$, which is not high enough for eliminating these noise contributions to the calculated property values.

In Figure~\ref{fig:pos_plot_mpil_di_r}{c}, a linear log-log correlation is observed for $\left(L/h_{\rm min}\right)_{\rm max}$. Points show a very small negative mean value and seem to be almost symmetrically distributed around it. This correlation shows a larger scatter in comparison to $L_{\rm tot}$ -- most points are distributed over almost an order of magnitude around the mean. Figure~\ref{fig:pos_plot_mpil_di_r}{d} shows a slightly different scenario to $L_{\rm tot}$: near-zero differences (green symbols) dominate for $\left(L/h_{\rm min}\right)_{\rm max}$ over all $\phi$ values, but some large-magnitude differences (predominantly blue symbols) also occur. There seems to be no obvious indication that non-zero values dominate over any longitudes, with the larger scatter attributed to this property being the combination of two independent AR properties. The largest differences between $B_{r}$- and $B_{\rm los}$-derived $\left(L/h_{\rm min}\right)_{\rm max}$ are on the order of hundreds of gauss and seem to occur mostly when using $B_{\rm los}$. Again, it is difficult to assess differences between $B_{\rm los}$ and $B_{r}$ as both $^{\star}$MPIL length and $h_{\rm min}$ above these MPILs vary between field components.

The scatter plot for $\log\left(R\right)$ in Figure~\ref{fig:pos_plot_mpil_di_r}{e} shows a reasonable linear correlation for data points from $|\phi| < 60\degree$ (black symbols). The linear correlation shifts above the unity diagonal for $60\degree \leqslant |\phi| < 75\degree$ (blue symbols) and more so for $|\phi| \geqslant 75\degree$ (red symbols), indicating that $B_{\rm los}$ gives systematically lower values compared to $B_{r}$ further from disk centre. All points are more or less equally distributed around the unity diagonal  with the smallest mean value of all properties and with most points distributed over $\approx$ 1.3 orders of magnitude. However, visualization of the data in Figure~\ref{fig:pos_plot_mpil_di_r}{f} shows that property differences for $|\phi| < 60\degree$ are a mixture of negative ($B_{\rm los}$ yielding larger $R$ values; blue symbols) and near-zero values (green symbols). This behavior, in locations where both property values should be essentially the same (disk centre), can be attributed to the different magnetic flux thresholds applied in determining $R$-value MPILs. For $|\phi| \geqslant 60\degree$, $\log\left[R\left(B_{r}\right)\right]$ becomes larger than $\log\left[R\left(B_{\rm los}\right)\right]$ (yellow\,--\,red symbols), with differences increasing at $|\phi| \geqslant 75\degree$. Since in this case, both flux thresholds used are above the typical noise levels (150 and 300 G for $B_{\rm los}$ and $B_{r}$, respectively), larger $B_{r}$ values with increasing $|\phi|$ must arise from a better definition of the MPILs, specially at far distances from the central meridian.

Figure~\ref{fig:pos_plot_alpha_beff_eising}{a} presents the scatter plot for Fourier spectral power index, $\alpha$, that has a mostly linear relationship over low values ({\it i.e.} $0.5-2.0$). Beyond that range, $\alpha\left(B_{r}\right)$ appears to saturate while $\alpha\left(B_{\rm los}\right)$ continues increasing, with the divergence from linearity and scatter increasing progressively with $|\phi|$ (blue and red symbols). The distribution of all differences is clearly  skewed towards $B_{\rm los}$ values with a mean value lying bellow the unity diagonal and negative large skewness. In this case, the 2$\sigma$ spread is only $\approx$ 0.6, since most points come from region with $|\phi| < 60\degree$. This location dependence is easily observed in Figure~\ref{fig:pos_plot_alpha_beff_eising}{b}, where most data points for $|\phi| < 45\degree$ have near-zero differences (green symbols) and those for $|\phi| > 45\degree$ are dominated by negative differences ($B_{\rm los}$ yielding larger $\alpha$ values; blue symbols). Although $B_{\rm los}$ and $B_{r}$ are equally affected by foreshortening effects that cut off small size scales at greater $|\phi|$, differences in the structure of magnetic features with differing sizes will affect $B_{\rm los}$ spatial-frequency sampling another way. While small-scale mostly-vertical flux tubes ({\it i.e.} network/plage fields) experience the usual diminishing of $B_{\rm los}$ field strength by viewing-angle, large-scale fields contained in AR sunspots will experience less diminishing of $B_{\rm los}$ in comparison to small-scale fields. This unbalanced decrease of LOS signal occurs because when viewed obliquely, large-scale (sunspot) fields include horizontal field strength components which are not detectable by HMI in the small-scale fields. Therefore this effect produces systematic diminishing of Fourier power at small scales ({\it i.e.} large $k$) more than that at large scales ({\it i.e.} small $k$), hence increasing the $\alpha$ calculated from $B_{\rm los}$.

For $B_{\rm eff}$, Figure~\ref{fig:pos_plot_alpha_beff_eising}{c} shows a generally linear log-log relation between $B_{\rm los}$- and $B_{r}$-derived values. Similar to $L_{\rm tot}$, the majority of data points (and therefore the difference mean) lie above the unity diagonal -- $B_{\rm eff}\left(B_{r}\right)$ is typically greater than $B_{\rm eff}\left(B_{\rm los}\right)$, as a consequence of the $B_{\rm los}$ field-strength underestimation. The distribution of all points around the unity diagonal shows a spread of 2$\sigma=$ 0.9, almost a order of magnitude. The correlation weakens for $60 \leqslant |\phi| < 75\degree$ (blue symbols) and practically disappears for $|\phi| \geqslant 75\degree$ (red symbols), but these data points still fall within the scatter for $|\phi| < 60\degree$ (black symbols). As shown in Figure~\ref{fig:pos_plot_alpha_beff_eising}{d}, $B_{r}$ yields values almost equal to or slightly larger than those from $B_{\rm los}$ for $|\phi| < 30\degree$ (yellowish-green symbols) with a higher frequency of large differences starting at $|\phi| \geqslant 30\degree$. As explained for the $L_{\rm tot}$ case, the flux threshold value of 100 Mx cm$^{-2}$ used for $B_{\rm eff}$ does not remove the effects of vector-field noise spatio-temporal patterns and therefore the asymmetric increasing difference with $|\phi|$ appears.

Finally, Figures~\ref{fig:pos_plot_alpha_beff_eising}{e} and \ref{fig:pos_plot_alpha_beff_eising}{f} present $E_{\rm Ising}$ as showing similar behaviour to $B_{\rm eff}$; a generally linear log-log relation with large scatter ($2\sigma \approx$ 1.15 orders of magnitude) and data points lying almost exclusively above the unity diagonal (positive mean). In contrast to $B_{\rm eff}$, there is a very wide range of values for $E_{\rm Ising}\left(B_{r}\right)$ over the smallest values of $E_{\rm Ising}\left(B_{\rm los}\right)$, that correspond to the secondary (low-frequency) peaks in Figure~\ref{fig:histograms}{f}. Values of $E_{\rm Ising}\left(B_{\rm los}\right) < 1.0$ result from SHARPs containing small ARs in the early stages of formation ({\it i.e.} a small number of quite separated opposite-polarity pixels). Correspondingly larger values of $E_{\rm Ising}\left(B_{r}\right)$ (observed at many longitudes) arise from the underestimation of field strength by $B_{\rm los}$. Although $E_{\rm ising}$ uses a pixel-consideration threshold of 100\,Mx\,cm$^{-2}$ -- lower than the maximum noise level in $B_{r}$ --  no longitudinal variation and/or east-west asymmetry is observed likely due to contributions from noisy pixels being few and small (opposite-polarity pixels separated by large distances).

On one hand, the LOS component is known to be less noisy ($\approx$ 5--10 G) but it also presents issues related to the viewing angle obliqueness: not only a field strength decrease but also, in extreme cases, the presence of false MPILs in complex sunspot ARs. The former issue is particularly clear in the behaviour of $\alpha$ with the longitudinal distance. The effect of false MPILs in MPIL-related properties could be difficult to observed since properties such as $L_{\rm tot}$ and  $R$ sum the contributions of all $^{\star}$MPIL segments and their contribution will be small in comparison to the real MPILs present in NOAA-numbered regions. On the other hand, the effect of noise patterns seen in $B_{r}$-calculated properties can be alleviated by choosing a high enough flux threshold or, alternatively, by performing statistical corrective analysis such as in \cite{2016ApJ...833L..31F}. However, using $B_{r}$ data should result in more consistent property representation with AR disk position.

No property in this study was observed to have significant variation with the HG latitudinal position. Although, the longitudinal variation of properties here found might not apply to ARs located at higher latitudes than those included in this study.

\subsection{Property-property Correlations}\label{ss:par_correl}

\begin{figure}
 \centerline{\includegraphics[width=1.1\textwidth,clip=]{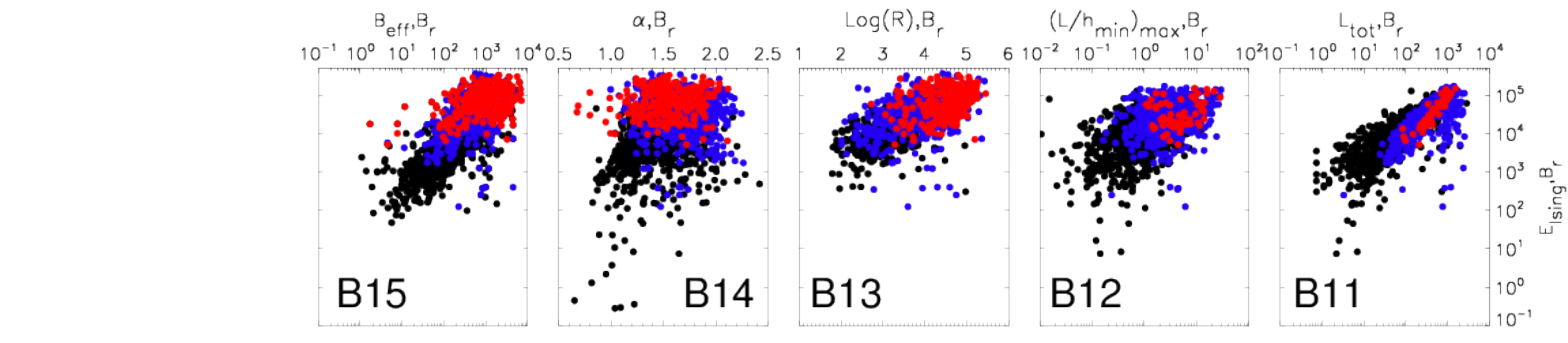}}
 \centerline{\includegraphics[width=1.1\textwidth,clip=]{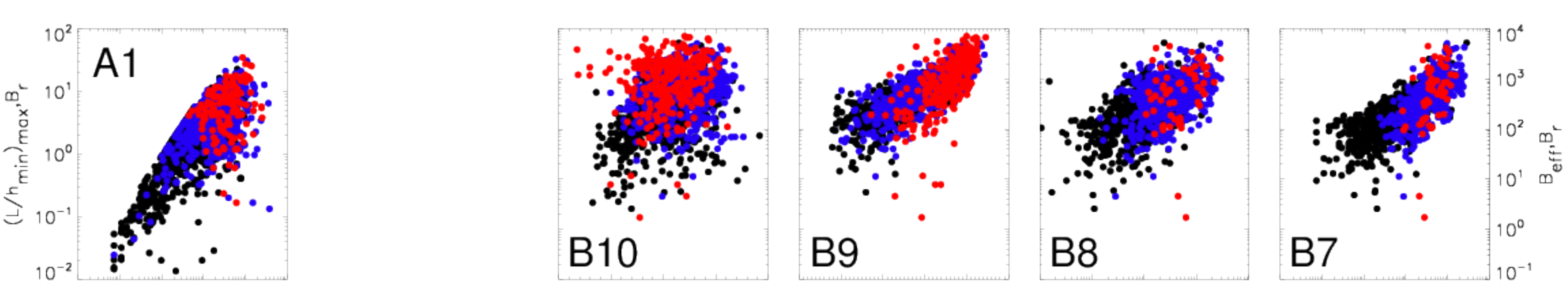}}
 \centerline{\includegraphics[width=1.1\textwidth,clip=]{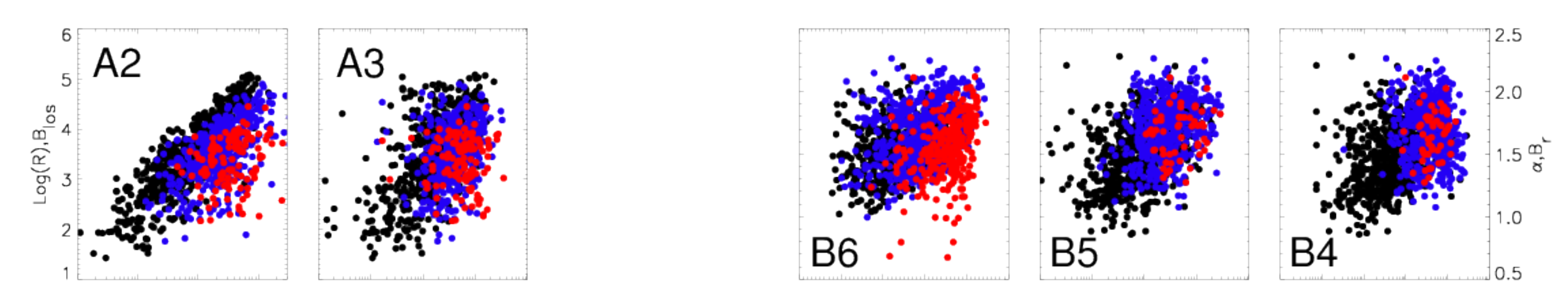}}
 \centerline{\includegraphics[width=1.1\textwidth,clip=]{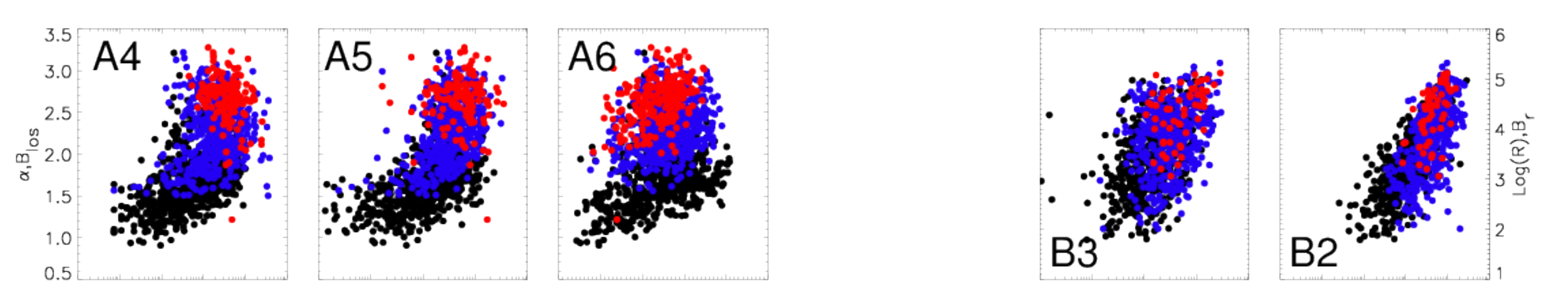}}
 \centerline{\includegraphics[width=1.1\textwidth,clip=]{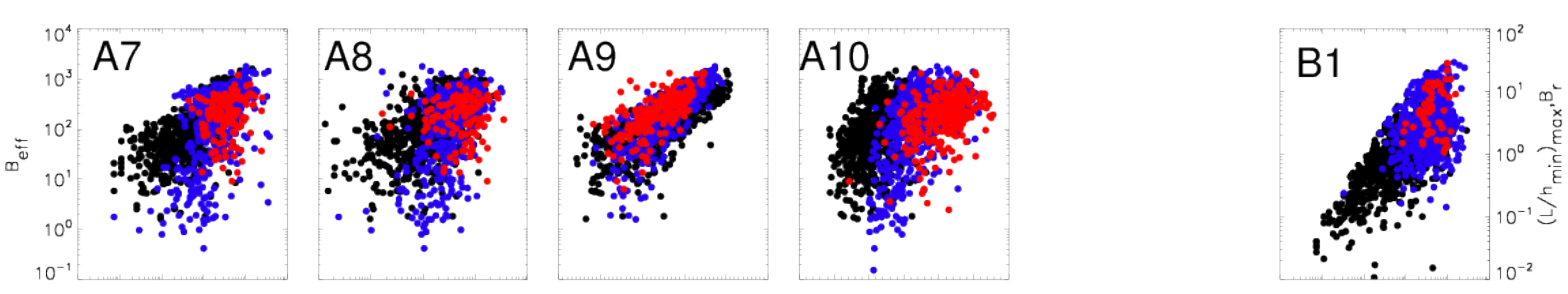}}
 \centerline{\includegraphics[width=1.1\textwidth,clip=]{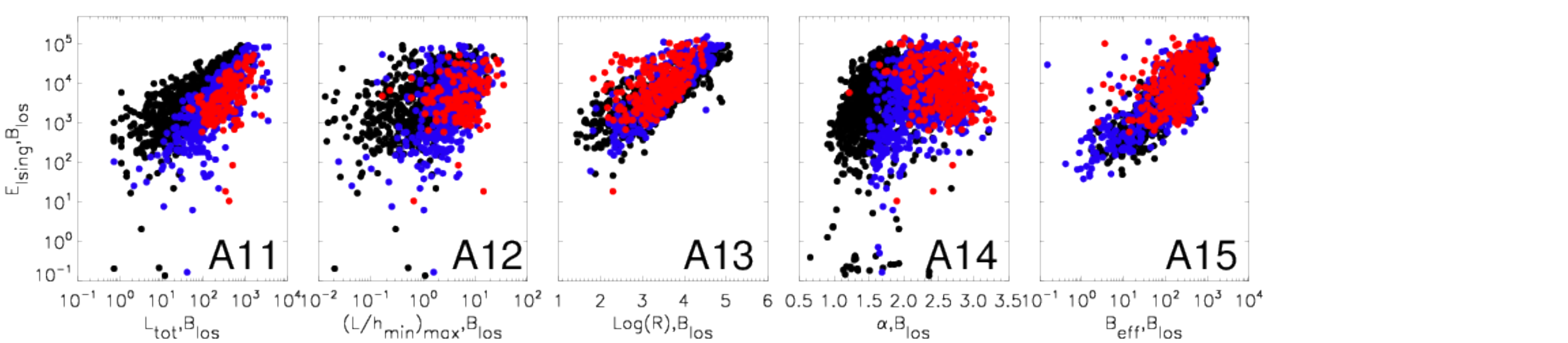}}
 \caption{Property-property plots for values calculated using $B_{\rm los}$ (A panels; below diagonal) and $B_{r}$ (B panels; above diagonal). All panels in a given row share a common abscissa (left for panels A, right for panels B), while all panels in a given column share a common ordinate (bottom for panels A, top for panels B). In the $B_{\rm los}$ case, from top to bottom (and left to right) properties are displayed in the order: $L_{\rm tot}$; $\left(L/h_{\rm min}\right)_{\rm max}$; $\log\left(R\right)$; $\alpha$; $B_{\rm eff}$; $E_{\rm Ising}$. For $B_{r}$, properties order is reverse as for $B_{\rm los}$. As in the left-hand panels of Figures~\ref{fig:pos_plot_mpil_di_r} and \ref{fig:pos_plot_alpha_beff_eising}, data points are colour-coded to represent three longitude groups: $|\phi| < 60\degree$ (black); $60\degree \leqslant |\phi| < 75\degree$ (blue); $|\phi| \geqslant 75\degree$ (red). In comparing the effect of $B_{\rm los}/B_{r}$ for same property pair, match the same number in panels A and B.}
 \label{fig:ppcorel}
 \end{figure}

In addition to exploring the behaviour between $B_{\rm los}$- and $B_{r}$-derived values of a given property, it is worth investigating correlations between different properties for a given field component. This is presented in Figure~\ref{fig:ppcorel} in the form of scatter plots for all pair-wise combinations of the six properties derived from $B_{\rm los}$ (panels A1-15) and $B_{r}$ (panels B1-15). Data points are colour-coded in the same way as those in the left-hand panels of Figures~\ref{fig:pos_plot_mpil_di_r} and \ref{fig:pos_plot_alpha_beff_eising}: $|\phi| < 60\degree$ (black); $60\degree \leqslant |\phi| < 75\degree$ (blue); $|\phi| \geqslant 75\degree$ (red). For practical purposes, a random 25\% of data points from $|\phi| < 60\degree$ are provided in each plot. Property self-correlation panels ({\it i.e.} on the diagonal) are omitted. Linear (Pearson) and nonlinear-rank (Spearman) correlation coefficients (CC) for each property pair are given in Tables~\ref{table2} and \ref{table3} of Appendix~\ref{s:ccorrel} for $B_{\rm los}$ and $B_{r}$ data, respectively. Uncertainty estimates for CC values are given in Figure~\ref{fig:correl_errors}.

Relationships between $B_{\rm los}$-derived properties in Figure~\ref{fig:ppcorel} (A panels) and Table~\ref{table2} span the full range of correlations from very weak to very strong. As expected, properties related to the same features correlate well -- {\it e.g.} correlations between the MPIL-related property pairs $L_{\rm tot}$-$\left(L/h_{\rm min}\right)_{\rm max}$ and $L_{\rm tot}$-$R$ are very strong ($|{\rm CC}| \geqslant 0.8$) and that for the magnetic-connectivity pair $B_{\rm eff}$-$E_{\rm Ising}$ is strong ($0.6 \leqslant |{\rm CC}| < 0.8$). However, good correlation is observed between pairs of different property types -- {\it e.g.} $B_{\rm eff}$ and $E_{\rm Ising}$ show at least strong correlation ($|{\rm CC}| \geqslant 0.6$) with $L_{\rm tot}$ and $R$. This is due to the magnetic-connectivity properties scaling with the amount and proximity of opposite-polarity flux; defining features of strong/long MPILs. CCs decrease with increasing $|\phi|$ for all property pairs, while the fractality-related property $\alpha$ shows at most moderate correlation ($|{\rm CC}| < 0.6$) with all properties.

Comparative $B_{r}$-derived property-property correlations are given in Figure~\ref{fig:ppcorel} (B panels) and Table~\ref{table3}, with results generally similar to the $B_{\rm los}$ case. Differences include most $B_{r}$-derived CCs being lower than those from $B_{\rm los}$ (although less clear in pairs with a magnetic-connectivity property) and the CC decrease with increasing $|\phi|$ being smaller in magnitude than for $B_{\rm los}$. The variation of linear and nonlinear CCs from $B_{\rm los}$ to $B_{r}$ can be understood in terms of more consistent property representation from $B_{r}$. As shown in Section~\ref{ss:var_pos}, using $B_{r}$ to calculate properties results in the reduction of viewing-angle bias in $B_{\rm los}$, such that CCs between some viewing-angle sensitive and insensitive property pairs increase for $|\phi| \geqslant 75\degree$ with $B_{r}$. This is expected for some property pairs as one property influences the calculation of the other -- {\it e.g.} $L_{\rm tot}$-$R$, where MPIL location (and hence length) is used to determine the spatial region within which flux is integrated into $R$. Without the common bias in $B_{\rm los}$, CCs between property pairs that are sensitive to viewing angle decrease for $|\phi| \geqslant 75\degree$ with $B_{r}$ ({\it e.g.} $R$-$B_{\rm eff}$, $R$-$E_{\rm Ising}$, and $B_{\rm eff}$-$E_{\rm Ising}$). 

From a total of 90 property-pair correlations (45 linear/45 nonlinear), 59 (29/30) decreased when $B_{\rm los}$ was replaced by $B_{r}$ in property calculations. Only 33 (13/20) of those correlations decreased by values larger than their associated standard errors (see Figure~\ref{fig:correl_errors} in Appendix for details). Such decreases in CC indicate that these properties may have a greater degree of independence than previously thought from $B_{\rm los}$ data. Although, only about 37\% of the studied properties showed significant decrease.

\subsection{Flaring Association}\label{ss:flaring}

\begin{figure}
 \centerline{\includegraphics[width=0.5\textwidth,clip=]{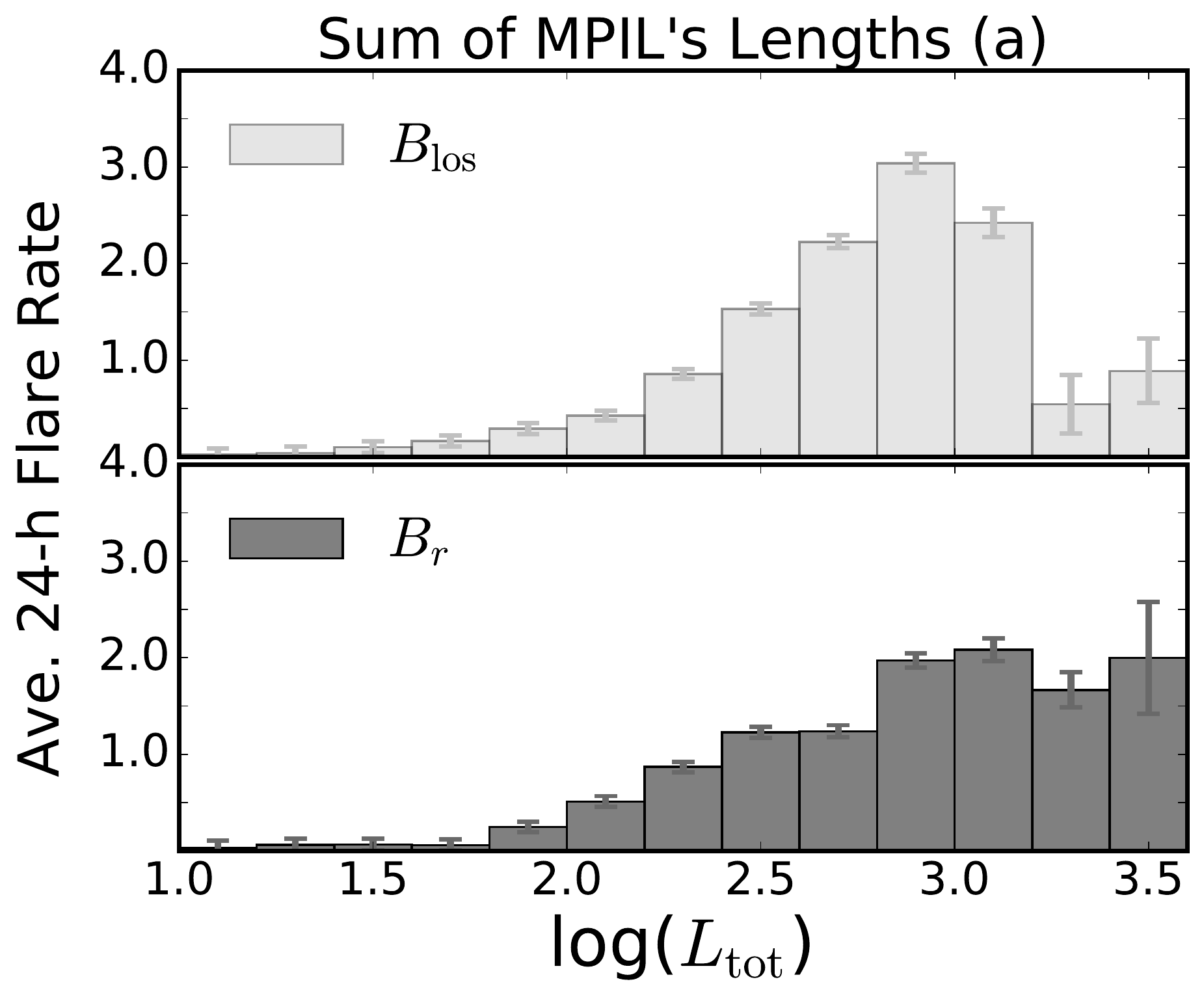}\includegraphics[width=0.5\textwidth,clip=]{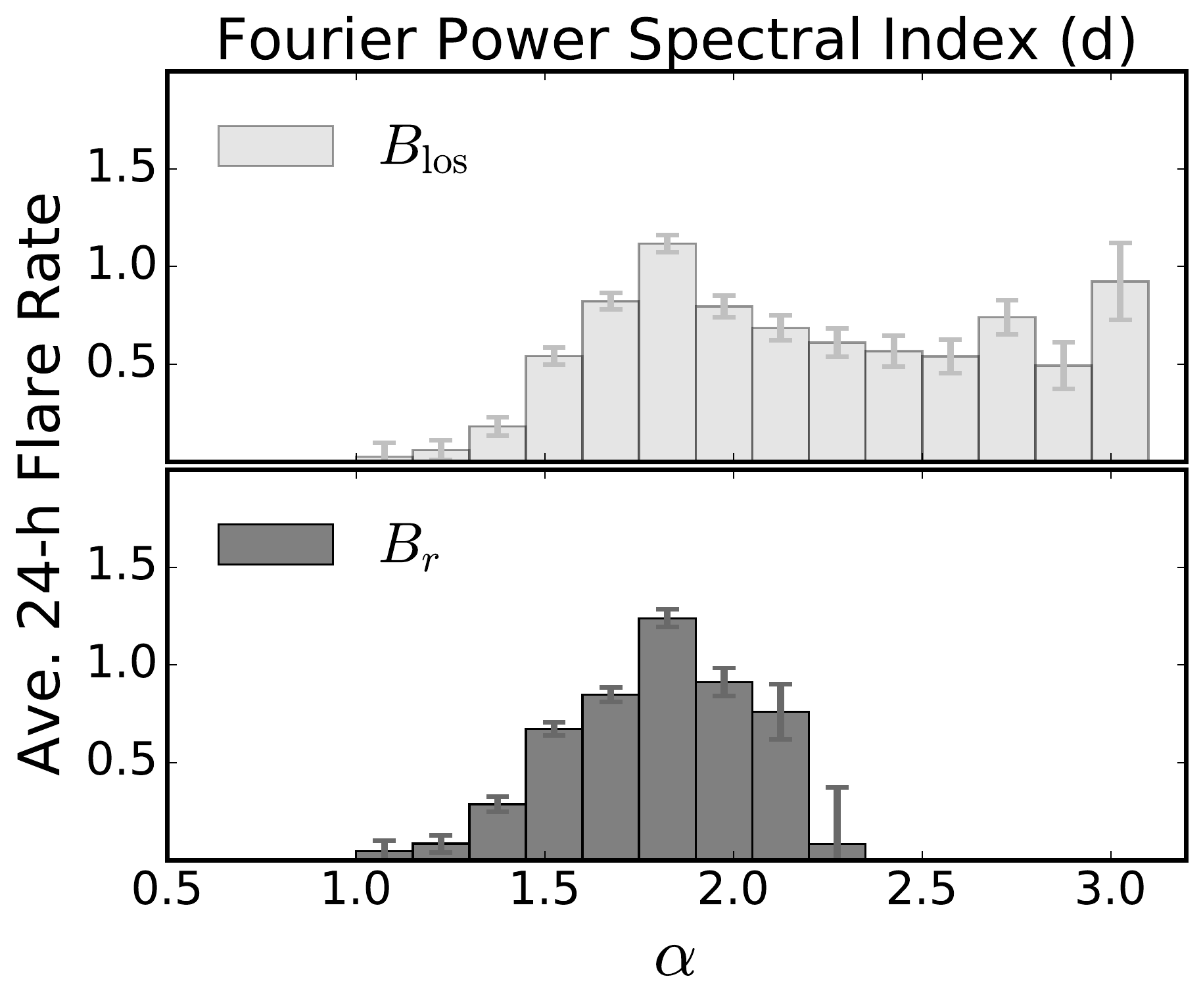}}
 \centerline{\includegraphics[width=0.5\textwidth,clip=]{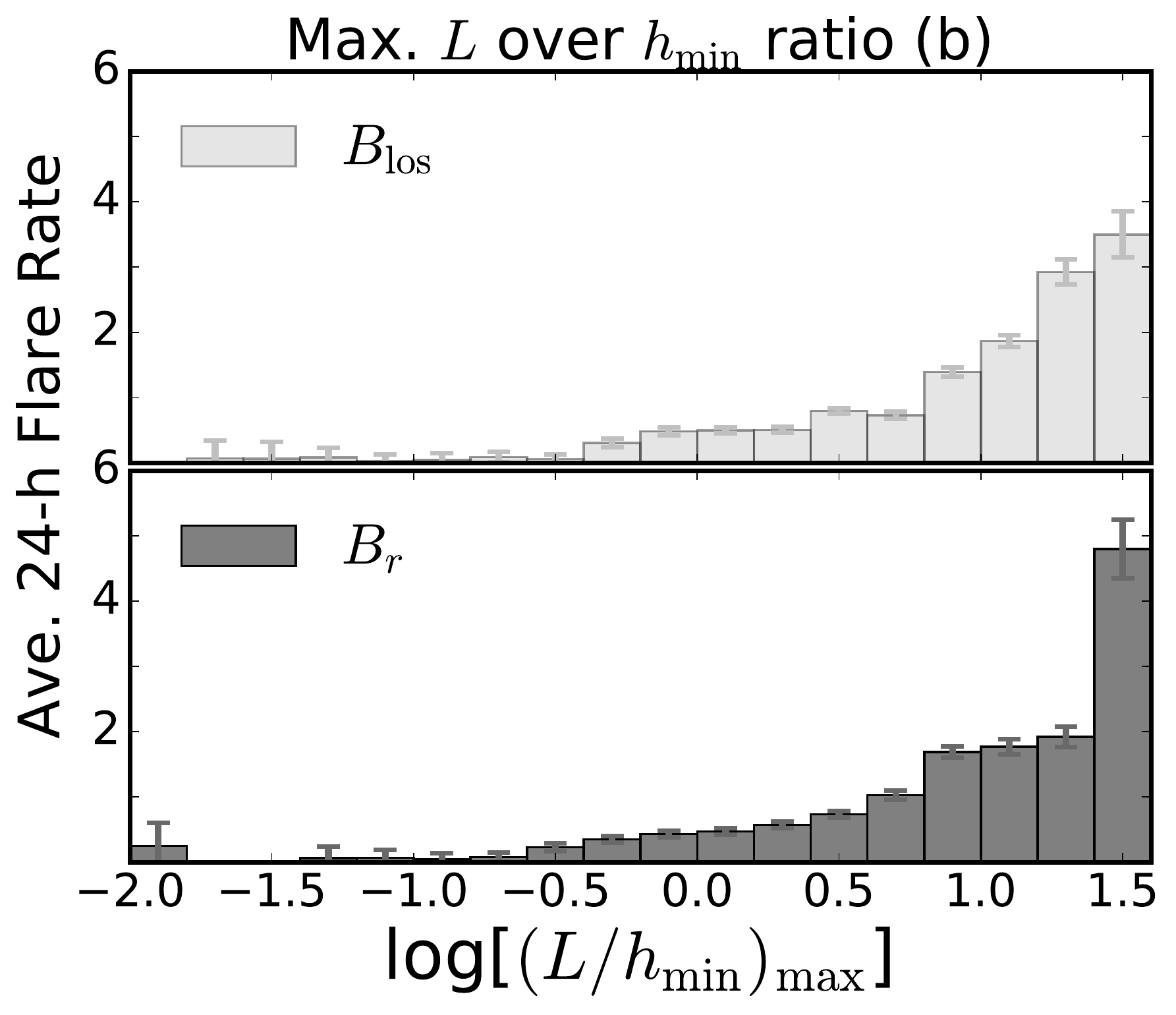}\includegraphics[width=0.5\textwidth,clip=]{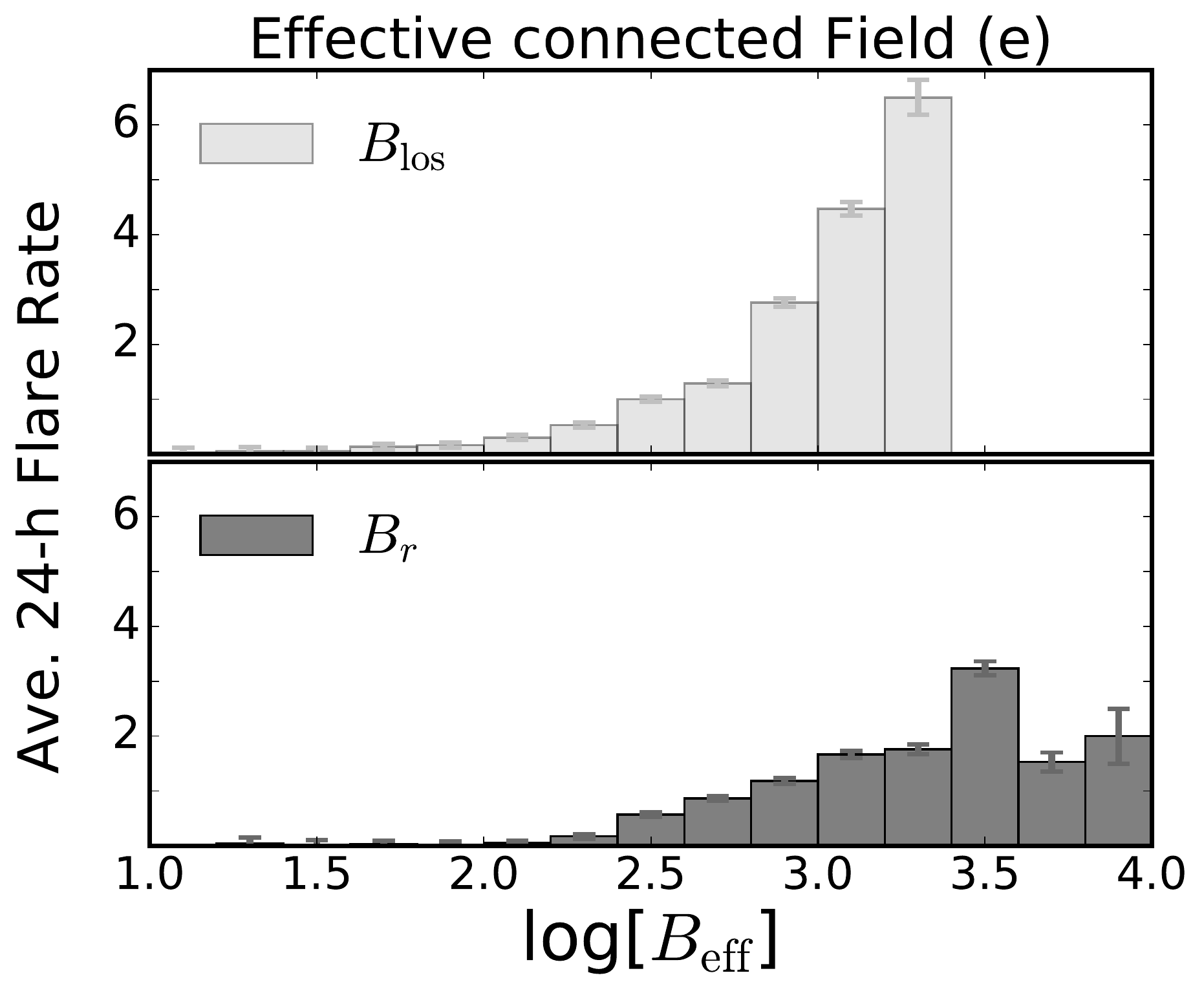}}
 \centerline{\includegraphics[width=0.5\textwidth,clip=]{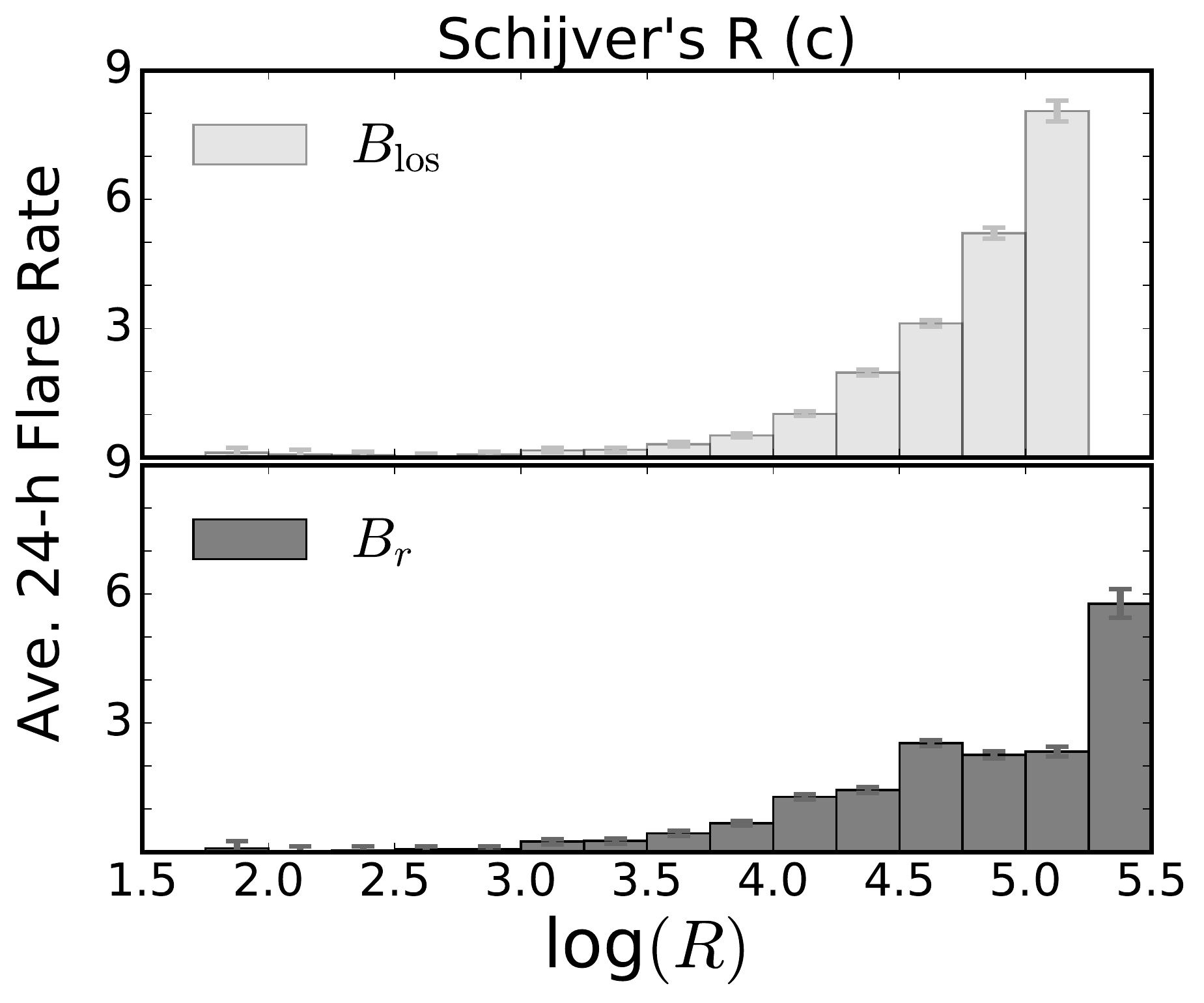}\includegraphics[width=0.5\textwidth,clip=]{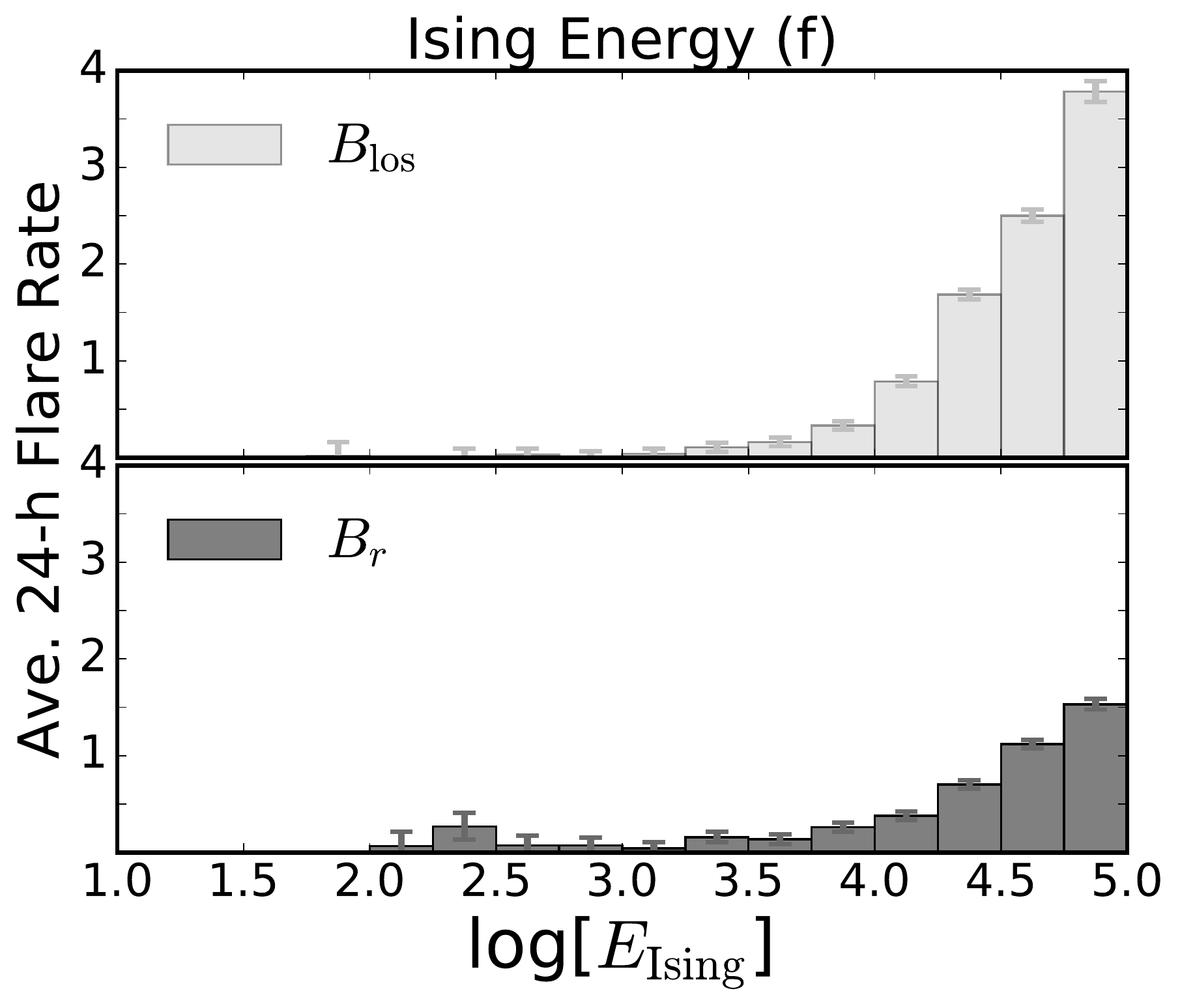}}
 \caption{Average C-class and greater flaring rates in the subsequent 24\,hr, with error bars indicating Poisson uncertainties ({\it i.e.} $N^{-1/2}$, where $N$ is number of SHARPs in a property-value bin). In each panel, grey-scale shading displays the flaring rates for properties calculated from $B_{\rm los}$ (light grey) and from $B_{r}$ (dark grey), while regions of overlap are indicated in mid-grey.}
 \label{fig:flaring_assoc_1}
 \end{figure}

For the purposes of flare forecasting it is necessary to know the relation between the calculated properties and observed flaring activity. As mentioned before, most of these properties have been reported as having useful association with occurrence of flares/CMEs in the AR they represent. Attention is paid here to the overall differences in flaring rates associated with $B_{\rm los}$- and $B_{r}$-derived properties. Figure~\ref{fig:flaring_assoc_1} presents average flaring rates observed in a 24-hr window after the property observation time. In each panel, rates are displayed for flares of C-class and greater for $B_{\rm los}$ (light grey, top plot), $B_{r}$ (dark grey, bottom plot), with error bars indicating Poisson uncertainties ({\it i.e.} $N^{-1/2}$, where $N$ is the number of HARPs in a property-value bin). It is very important to note that $B_{\rm los}$- and $B_{r}$-derived flaring rates for the same property-value bin do not relate to the same HARPs, due to the differences between $B_{\rm los}$ and $B_{r}$ property values presented in the scatter-plots of Figures~\ref{fig:pos_plot_mpil_di_r} and \ref{fig:pos_plot_alpha_beff_eising}.

The general behaviour of most panels in Figure~\ref{fig:flaring_assoc_1} ({\it i.e.} except panel d that represents $\alpha$) is that average flaring rates increase with increasing property value. This tendency is a good indicator that those properties might capture good statistical association of future flaring activity levels in ARs. In addition, although properties like $R$ and $B_{\rm eff}$ span different dynamic ranges between $B_{r}$ and $B_{\rm los}$ values, $B_{r}$ distributions appear to show a different ({\it i.e.} slower) rate in increase in comparison to the $B_{\rm los}$ distributions. The difference between $B_{r}$- and $B_{\rm los}$-associated flaring rates is due to many of the HARPs being corrected upwards into the higher-valued property bins. Distinctly different behaviour is displayed by $\alpha$ in Figure~\ref{fig:flaring_assoc_1}{d}, where restriction of values to $\alpha < 2.5$ presented in Figures~\ref{fig:histograms}{d} and \ref{fig:pos_plot_alpha_beff_eising}{a,b} is again evident. Folding of $B_{\rm los}$-derived values into the smaller $B_{r}$ range yields a slight increase in most $B_{r}$ flaring rates in that range, enhancing the peak around $\alpha = 5/3 \approx 1.67$. However, the maximum rate achieved by $\alpha$ is lower than all other properties by at least a factor of 2. This implies that there is a much greater mixture of flaring and non-flaring populations in all $\alpha$ property-value bins, indicating that, although better determine by $B_{r}$, it is less useful as flare predictor than other properties here studied. In order to determine the forecasting advantage of using $B_{r}$ data over $B_{\rm los}$, proper forecasting and verification analysis must be performed, which lies beyond the scope of this investigation.

\section{Conclusions}\label{s:concl}

This work presents a statistical study of differences in AR properties calculated from different components of the magnetic field, focusing on six properties that have been claimed to quantify AR potential to produce flares/CMEs: total length of strong MPILs, $L_{\rm tot}$; maximum ratio of strong MPIL length to minimum height of critical decay index, $\left(L/h_{\rm min}\right)_{\rm max}$; Schrijver's $R$ value; Fourier spectral power index, $\alpha$; effective connected magnetic field strength, $B_{\rm eff}$; Ising energy, $E_{\rm Ising}$. All six properties were calculated from SDO/HMI SHARP CEA NRT LOS ($B_{\rm los}$) and spherical-radial ($B_{r}$) magnetograms, with the dependence on longitudinal position, inter-property correlations, and associated flaring rates studied.

Overall, property-value distributions calculated from $B_{\rm los}$ and $B_{r}$ are similar. Differences in observed ranges indicate that $B_{r}$ typically yields marginally larger values than $B_{\rm los}$ for all properties except $\left(L/h_{\rm min}\right)_{\rm max}$ (that shows a scatter of values from $B_{\rm los}$ sometimes larger than those from $B_{r}$) and $\alpha$ (that shows a constraining of values from $B_{r}$ to $< 2.5$, compared to those from $B_{\rm los}$ spreading out to $3.3$). Properties such as $L_{\rm tot}$, $\left(L/h_{\rm min}\right)_{\rm max}$, and $E_{\rm Ising}$ do not show a strong dependence on longitudinal position, while $\log\left(R\right)$ ($B_{r}$ greater), $\alpha$ ($B_{\rm los}$ greater), and $B_{\rm eff}$ ($B_{r}$ greater) display progressively increasing differences between $B_{\rm los}$ and $B_{r}$ with increasing absolute longitude. Although $B_{r}$ data contains higher levels of noise with complex spatio-temporal patterns, considering the difficulty that the LOS component has in estimating the normal-to-surface field at increasing longitudinal distance from central meridian, and the systematic field strength underestimation, the results here presented support the conclusion that $B_{r}$-derived properties are a more consistent representation of AR properties.

The properties considered here show a wide range of correlations, with linear (Pearson) and nonlinear-rank (Spearman) correlation coefficients similar in value for all cases. All property-pair correlations become weaker with increasing longitudinal distance from central meridian, with decreases for $B_{\rm los}$ more pronounced than those for $B_{r}$). While most property pairs show lower correlation levels from $B_{r}$ than from $B_{\rm los}$, only approximately a third of them show significantly lower values when compared to their associated errors. This implies that some properties calculated from $B_{r}$ data may provide more independent information about the physical state of the AR photospheric magnetic field than the same properties calculated from $B_{\rm los}$.

Binned property-value 24-hr flaring rates from $B_{\rm los}$ and $B_{r}$ differ, due to each SHARP resulting in different values for a given property. However, the differences in flaring rate distributions between these two forms of magnetogram data are consistent with the redistribution of properties to values that are typically larger for $B_{r}$ than for $B_{\rm los}$ (except $\alpha$, which is folded into a more constrained property range). Notably, flaring rates for most properties increase with increasing property values, apart from $\alpha$ where they peak for values close to $5/3$.

Although $B_{r}$ data seem to provide more consistent determination of AR properties across the solar disk, there is an additional aspect of these data to consider. While it appears that $B_{r}$ should be preferentially considered over $B_{\rm los}$ for locations with significant viewing-angle bias, the less noisy $B_{\rm los}$ data may be more suitable towards disk centre. However, to determine the real utility of $B_{r}$ data for flare forecasting, AR properties calculated from both field components should be compared using a variety of forecast methods ({\it e.g.} discriminant analysis, machine learning, {\it etc}.) Definitive statements on the forecasting advantage of using $B_{r}$-derived properties over those from $B_{\rm los}$ can only then be made after forecast verification.

%
\begin{acks}
This research was funded by the European Union Horizon 2020 research and innovation programme under grant agreement No.~640216 (FLARECAST). SHARP data were provided courtesy of NASA/SDO and the HMI science team, and hosted for FLARECAST (\url{http://flarecast.eu}) by the MEDOC data and operations centre (\url{http://medoc.ias.u-psud.fr}; CNES/CNRS/Univ. Paris-Sud). We thank the anonymous referee for their comments; they certainly helped improving the manuscript.
\end{acks}

\section*{Disclosure of Potential Conflicts of Interest}
The authors declare that they have no conflicts of interest.

%
%
\bibliographystyle{spr-mp-sola}
\bibliography{ref.bib}

%
\appendix   

\section{Correlation Coefficients}\label{s:ccorrel}

\begin{figure}
 \centerline{\includegraphics[width=0.6\textwidth,clip=]{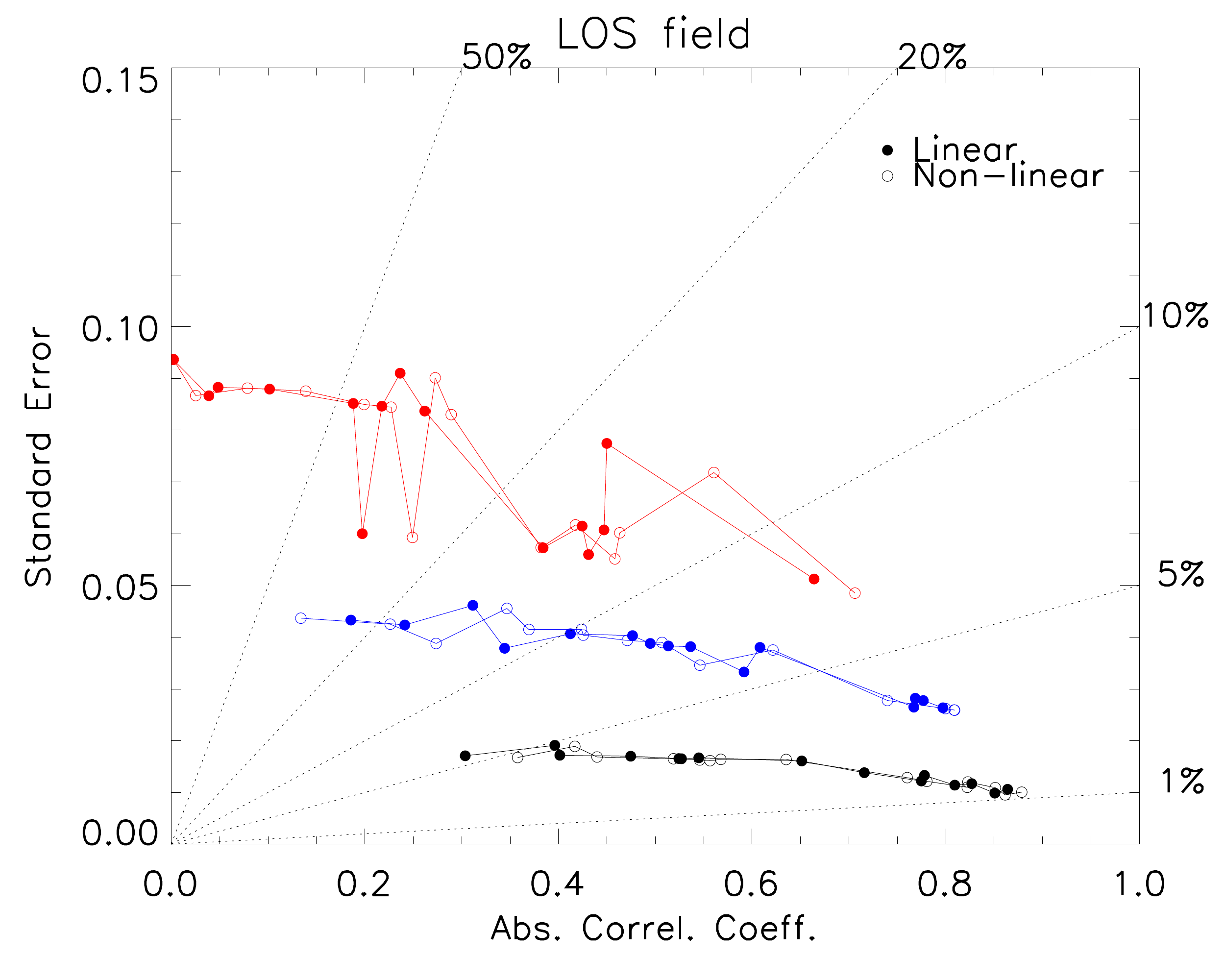}\includegraphics[width=0.6\textwidth,clip=]{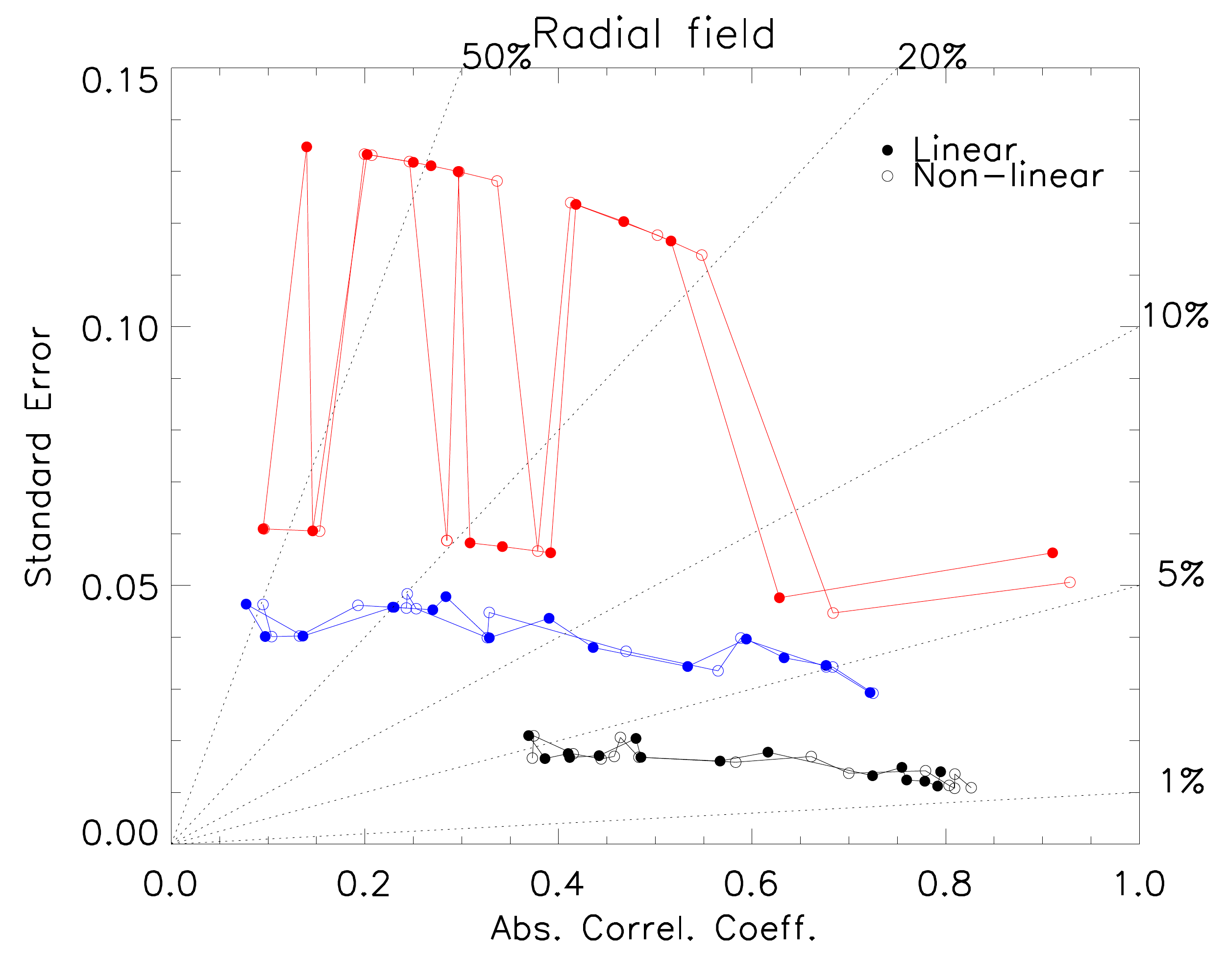}}
 \caption{Error analysis for correlation coefficients displayed in Tables~\ref{table2} and \ref{table3}. Left and right panels corresponds to property-pair (linear and nonlinear) correlations calculated from $B_{\rm los}$ and $B_{r}$, correspondingly. Color code corresponds to the same used in Figure \ref{fig:ppcorel}.}
 \label{fig:correl_errors}
 \end{figure}

Tables~\ref{table2} and \ref{table3} each contain values of linear (Pearson) and nonlinear-rank (Spearman) correlation coefficients (CC) for all AR property-pair combinations for $B_{\rm los}$ and $B_{r}$, respectively. Coefficients are calculated separately for the three groups of SHARP longitudes with data points colour-coded in Figures~\ref{fig:ppcorel} and \ref{fig:ppcorel}: $|\phi| < 60\degree$ (black); $60 \leqslant |\phi| < 75\degree$ (blue); $|\phi| \geqslant 75\degree$ (red). Figure \ref{fig:correl_errors} presents the errors for correlation coefficients contained in Tables~\ref{table2} and \ref{table3}. In both panels, plots of standard error {\it versus} CC value for linear (filled circles) and non-linear (open circles) correlations in all three longitudinal-position groups (same colour code applies). Left panel correspond to $B_{\rm los}$-calculated properties and right panel corresponds to $B_{r}$-calculated properties. Standard error for a (linear/non-linear) correlation coefficient is estimated as

\begin{equation}
{\rm SE}_{r} = \sqrt{\frac{1-r^{2}}{n-2}};
\end{equation}

\noindent
where $r$ can be Pearson or Spearman CC, and $n$ is the number of points used in their calculations. Dotted lines in Figure \ref{fig:correl_errors} plots mark ranges for expressing the standard errors in percentage of the CC value.

\begin{landscape}
\begin{table}
\caption{Linear (Pearson) and nonlinear-rank (Spearman) correlation coefficients for $B_{\rm los}$-derived properties.}
\setlength\tabcolsep{2.5pt}
\begin{tabular}{lccccccccccc}%
\hline
HG                                 & Ordinate & \multicolumn{10}{c}{Abscissa property} \\
longitude                          & property & \multicolumn{5}{c}{Linear correlation coefficient} & \multicolumn{5}{c}{Nonlinear-rank correlation coefficient} \\
$\left[\degree\right]$             &          & $\log\left[\left(\frac{L}{h_{\rm min}}\right)_{\rm max}\right]$ & $\log\left(R\right)$ & $\alpha$ & $\log\left(B_{\rm eff}\right)$ & $\log\left(E_{\rm Ising}\right)$ & $\log\left[\left(\frac{L}{h_{\rm min}}\right)_{\rm max}\right]$ & $\log\left(R\right)$ & $\alpha$ & $\log\left(B_{\rm eff}\right)$ & $\log\left(E_{\rm Ising}\right)$ \\
\hline
$|\phi| < 60$\dotfill              & $\log\left(L_{\rm tot}\right)$                                  &  $0.809$ &  $0.864$ &  $0.475$ &  $0.716$ &  $0.775$ &  $0.780$ &  $0.879$ &  $0.519$ &  $0.760$ &  $0.822$ \\
\dotfill                           & $\log\left[\left(\frac{L}{h_{\rm min}}\right)_{\rm max}\right]$ &  \ldots  &  $0.651$ &  $0.527$ &  $0.545$ &  $0.524$ &  \ldots  &  $0.635$ &  $0.556$ &  $0.568$ &  $0.546$ \\
\dotfill                           & $\log\left(R\right)$                                            &  \ldots  &  \ldots  &  $0.396$ &  $0.778$ &  $0.827$ &  \ldots  &  \ldots  &  $0.417$ &  $0.823$ &  $0.851$ \\
\dotfill                           & $\alpha$                                                        &  \ldots  &  \ldots  &  \ldots  &  $0.401$ &  $0.304$ &  \ldots  &  \ldots  &  \ldots  &  $0.440$ &  $0.358$ \\
\dotfill                           & $\log\left(B_{\rm eff}\right)$                                  &  \ldots  &  \ldots  &  \ldots  &  \ldots  &  $0.851$ &  \ldots  &  \ldots  &  \ldots  &  \ldots  &  $0.861$ \\
\hline
$60 \leqslant |\phi| < 75$\dotfill & $\log\left(L_{\rm tot}\right)$                                  &  $0.513$ &  $0.608$ &  $0.185$ &  $0.536$ &  $0.767$ &  $0.425$ &  $0.622$ &  $0.134$ &  $0.507$ &  $0.809$ \\
\dotfill                           & $\log\left[\left(\frac{L}{h_{\rm min}}\right)_{\rm max}\right]$ &  \ldots  &  $0.312$ &  $0.412$ &  $0.476$ &  $0.495$ &  \ldots  &  $0.347$ &  $0.369$ &  $0.424$ &  $0.471$ \\
\dotfill                           & $\log\left(R\right)$                                            &  \ldots  &  \ldots  &  $0.241$ &  $0.777$ &  $0.797$ &  \ldots  &  \ldots  &  $0.226$ &  $0.809$ &  $0.799$ \\
\dotfill                           & $\alpha$                                                        &  \ldots  &  \ldots  &  \ldots  &  $0.592$ &  $0.344$ &  \ldots  &  \ldots  &  \ldots  &  $0.546$ &  $0.274$ \\
\dotfill                           & $\log\left(B_{\rm eff}\right)$                                  &  \ldots  &  \ldots  &  \ldots  &  \ldots  &  $0.776$ &  \ldots  &  \ldots  &  \ldots  &  \ldots  &  $0.739$ \\
\hline
$|\phi| \geqslant 75$\dotfill      & $\log\left(L_{\rm tot}\right)$                                  &  $0.217$ &  $0.236$ & $-0.262$ &  $0.102$ &  $0.450$ &  $0.289$ &  $0.273$ & $-0.227$ &  $0.139$ &  $0.560$ \\
\dotfill                           & $\log\left[\left(\frac{L}{h_{\rm min}}\right)_{\rm max}\right]$ &  \ldots  &  $0.002$ & $-0.039$ &  $0.048$ &  $0.188$ &  \ldots  &  $0.002$ & $-0.025$ &  $0.078$ &  $0.199$ \\
\dotfill                           & $\log\left(R\right)$                                            &  \ldots  &  \ldots  &  $0.424$ &  $0.663$ &  $0.446$ &  \ldots  &  \ldots  &  $0.417$ &  $0.706$ &  $0.463$ \\
\dotfill                           & $\alpha$                                                        &  \ldots  &  \ldots  &  \ldots  &  $0.384$ & $-0.197$ &  \ldots  &  \ldots  &  \ldots  &  $0.382$ & $-0.249$ \\
\dotfill                           & $\log\left(B_{\rm eff}\right)$                                  &  \ldots  &  \ldots  &  \ldots  &  \ldots  &  $0.431$ &  \ldots  &  \ldots  &  \ldots  &  \ldots  &  $0.458$ \\
\hline
\end{tabular}
\label{table2}
\end{table}
\end{landscape}

\begin{landscape}
\begin{table}
\caption{Linear (Pearson) and nonlinear-rank (Spearman) correlation coefficients for $B_{r}$-derived properties.}
\setlength\tabcolsep{2.5pt}
\begin{tabular}{lccccccccccc}%
\hline
HG                                 & Ordinate & \multicolumn{10}{c}{Abscissa property} \\
longitude                          & property & \multicolumn{5}{c}{Linear correlation coefficient} & \multicolumn{5}{c}{Nonlinear-rank correlation coefficient} \\
$\left[\degree\right]$             &          & $\log\left[\left(\frac{L}{h_{\rm min}}\right)_{\rm max}\right]$ & $\log\left(R\right)$ & $\alpha$ & $\log\left(B_{\rm eff}\right)$ & $\log\left(E_{\rm Ising}\right)$ & $\log\left[\left(\frac{L}{h_{\rm min}}\right)_{\rm max}\right]$ & $\log\left(R\right)$ & $\alpha$ & $\log\left(B_{\rm eff}\right)$ & $\log\left(E_{\rm Ising}\right)$ \\
\hline
$|\phi| < 60$\dotfill              & $\log\left(L_{\rm tot}\right)$                                  &  $0.724$ &  $0.795$ &  $0.442$ &  $0.778$ &  $0.760$ &  $0.700$ &  $0.809$ &  $0.458$ &  $0.826$ &  $0.803$ \\
\dotfill                           & $\log\left[\left(\frac{L}{h_{\rm min}}\right)_{\rm max}\right]$ &  \ldots  &  $0.480$ &  $0.410$ &  $0.567$ &  $0.485$ &  \ldots  &  $0.464$ &  $0.415$ &  $0.583$ &  $0.483$ \\
\dotfill                           & $\log\left(R\right)$                                            &  \ldots  &  \ldots  &  $0.369$ &  $0.755$ &  $0.616$ &  \ldots  &  \ldots  &  $0.374$ &  $0.779$ &  $0.661$ \\
\dotfill                           & $\alpha$                                                        &  \ldots  &  \ldots  &  \ldots  &  $0.412$ &  $0.386$ &  \ldots  &  \ldots  &  \ldots  &  $0.444$ &  $0.373$ \\
\dotfill                           & $\log\left(B_{\rm eff}\right)$                                  &  \ldots  &  \ldots  &  \ldots  &  \ldots  &  $0.792$ &  \ldots  &  \ldots  &  \ldots  &  \ldots  &  $0.809$ \\
\hline
$60 \leqslant |\phi| < 75$\dotfill & $\log\left(L_{\rm tot}\right)$                                  &  $0.230$ &  $0.594$ & $-0.008$ &  $0.676$ &  $0.633$ &  $0.193$ &  $0.588$ & $-0.095$ &  $0.683$ &  $0.677$ \\
\dotfill                           & $\log\left[\left(\frac{L}{h_{\rm min}}\right)_{\rm max}\right]$ &  \ldots  &  $0.283$ &  $0.270$ &  $0.390$ &  $0.229$ &  \ldots  &  $0.244$ &  $0.253$ &  $0.328$ &  $0.243$ \\
\dotfill                           & $\log\left(R\right)$                                            &  \ldots  &  \ldots  &  $0.329$ &  $0.722$ & $0.436$ &  \ldots  &  \ldots  &  $0.326$ &  $0.725$ &  $0.470$ \\
\dotfill                           & $\alpha$                                                        &  \ldots  &  \ldots  &  \ldots  &  $0.136$ & $-0.097$ &  \ldots  &  \ldots  &  \ldots  &  $0.132$ & $-0.103$ \\
\dotfill                           & $\log\left(B_{\rm eff}\right)$                                  &  \ldots  &  \ldots  &  \ldots  &  \ldots  &  $0.533$ &  \ldots  &  \ldots  &  \ldots  &  \ldots  &  $0.564$ \\
\hline
$|\phi| \geqslant 75$\dotfill      & $\log\left(L_{\rm tot}\right)$                                  &  $0.268$ &  $0.516$ &  $0.140$ &  $0.467$ &  $0.910$ &  $0.246$ &  $0.548$ &  $0.200$ &  $0.502$ &  $0.928$ \\
\dotfill                           & $\log\left[\left(\frac{L}{h_{\rm min}}\right)_{\rm max}\right]$ &  \ldots  &  $0.418$ &  $0.202$ &  $0.250$ &  $0.296$ &  \ldots  &  $0.413$ &  $0.207$ &  $0.337$ &  $0.297$ \\
\dotfill                           & $\log\left(R\right)$                                            &  \ldots  &  \ldots  &  $0.342$ &  $0.628$ &  $0.309$ &  \ldots  &  \ldots  &  $0.379$ &  $0.684$ &  $0.285$ \\
\dotfill                           & $\alpha$                                                        &  \ldots  &  \ldots  &  \ldots  &  $0.146$ &  $0.095$ &  \ldots  &  \ldots  &  \ldots  &  $0.153$ &  $0.096$ \\
\dotfill                           & $\log\left(B_{\rm eff}\right)$                                  &  \ldots  &  \ldots  &  \ldots  &  \ldots  &  $0.392$ &  \ldots  &  \ldots  &  \ldots  &  \ldots  &  $0.285$ \\
\hline
\end{tabular}
\label{table3}
\end{table}
\end{landscape}

\end{article} 
\end{document}